\documentclass[jpg,superscriptaddress,12pt]{iopart}

\usepackage{iopams} 
\usepackage{graphicx}
\usepackage[pdftex]{hyperref}
\usepackage{lineno}

\begin{document}

\title[Low-energy enhancement and fluctuations of $\gamma$-ray strength functions in $^{56,57}$Fe]{Low-energy enhancement and fluctuations of $\gamma$-ray strength functions in $^{56,57}$Fe: test of the Brink-Axel hypothesis}

\author{A.~C.~Larsen$^1$, M.~Guttormsen$^1$, N.~Blasi$^2$, A.~Bracco$^{2,3}$, F.~Camera$^{2,3}$, 
L.~Crespo~Campo$^1$, T.~K.~Eriksen$^{1,4}$,  A.~G\"{o}rgen$^1$, T.~W.~Hagen$^1$, V.~W.~Ingeberg$^1$, 
B.~V.~Kheswa$^1$, S.~Leoni$^{2,3}$, 
J.~E.~Midtb\o$^1$, B.~Million$^2$, H.~T.~Nyhus$^1$, T.~Renstr{\o}m$^1$, S.~J.~Rose$^1$, I.~E.~Ruud$^1$, 
S.~Siem$^1$, T.~G.~Tornyi$^{1,4}$, G.~M.~Tveten$^1$, A.~V.~Voinov$^5$, M.~Wiedeking$^6$, and F.~Zeiser$^1$}
\ead{a.c.larsen@fys.uio.no}
\address{$^1$ Department of Physics, University of Oslo, N-0316 Oslo, Norway}

\address{$^2$ INFN, Sezione di Milano, Milano, Italy}
\address{$^3$ Dipartimento di Fisica, University of Milano, Milano, Italy}
\address{$^4$ Department of Nuclear Physics, Australian National University, Canberra, Australia}
\address{$^5$ Department of Physics and Astronomy, Ohio University, Athens, Ohio 45701, USA}
\address{$^6$ iThemba LABS, P.O.  Box 722, 7129 Somerset West, South Africa}

\vspace{10pt}
\begin{indented}
\item[]January 2017
\end{indented}

\begin{abstract}
Nuclear level densities and $\gamma$-ray strength functions of $^{56,57}$Fe have been extracted 
from proton-$\gamma$ coincidences. 
A low-energy enhancement in the $\gamma$-ray strength functions up to a factor 
of 30 over common theoretical E1 models is confirmed.
Angular distributions of the low-energy enhancement in $^{57}$Fe
indicate its dipole nature, in agreement with findings for $^{56}$Fe. 
The high statistics and the excellent energy resolution of
the large-volume LaBr$_{3}$(Ce) detectors allowed for a thorough analysis of  
$\gamma$ strength as function of excitation energy. 
Taking into account the presence of strong Porter-Thomas fluctuations,
there is no indication of any 
significant excitation-energy dependence in the $\gamma$-ray strength function, 
in support of the generalized Brink-Axel hypothesis. 
\end{abstract}

%
\vspace{2pc}
%
\submitto{\JPG}
%
%
%

\section{Introduction}
\label{sec:intro}

One of the long-standing experimental and theoretical challenges within nuclear physics is the 
determination of the nucleus' available quantum levels and the decay properties of these levels 
in the excitation-energy region between the ground state and up to the particle threshold(s). 
In this intermediate excitation-energy region, often called the \textit{quasicontinuum},
the \textit{nuclear level density} (NLD)  and the average, reduced $\gamma$-decay probability, i.e. the 
\textit{$\gamma$-strength function} ($\gamma$SF), shed light on the dynamic behavior of the nucleus. 
Apart from 
providing information on basic nuclear properties, these quantities are also indispensable input
for calculations of, e.g., neutron-capture cross sections. These cross sections are 
of great importance for applications 
such as the astrophysical heavy-element nucleosynthesis~\cite{arnould2003,arnould2007} 
and modeling of next-generation nuclear
power plants~\cite{aliberti2004,aliberti2006}. 

Amongst a handful of experimental techniques, the Oslo method~\cite{schiller2000} has been established 
as one of the promising approaches to obtain experimental information on the NLD and $\gamma$SF. The
advantage of the Oslo method compared to other techniques 
is that both these quantities can be extracted from one and the same experiment,
utilizing typically a charged-particle reaction to record particle-$\gamma$ coincidences, in which 
the structural shape of the NLD and the $\gamma$SF can be determined. By measuring the energy of the 
outgoing charged particle, the initial excitation energy of the residual nucleus is determined.  
The $\gamma$ rays de-exciting this initial excitation energy are recorded in coincidence, thus obtaining
$\gamma$ spectra as function of initial excitation-energy. 

In 2004, an unexpected enhancement of the $\gamma$SF for low 
transition energies ($E_\gamma \lesssim 3$ MeV) was discovered in the iron isotopes $^{56,57}$Fe~\cite{voinov2004}. 
This feature was not predicted by any theoretically derived $\gamma$SFs; 
in fact,
the $\gamma$SF data showed an enhancement of more than a factor of 10 compared to typical 
models for the E1 strength~\cite{voinov2004}. In the following years 
this enhancement, also called \textit{upbend}, 
was found in many medium-mass nuclei, including $^{43-45}$Sc~\cite{larsen2007,burger2012},
$^{60}$Ni~\cite{voinov2010},
$^{73,74}$Ge~\cite{renstrom2016}, and Mo isotopes~\cite{guttormsen2005,wiedeking2012,tveten2016}.
To date, the heaviest nuclei where the upbend has been seen are $^{138,139}$La~\cite{kheswa2015} and
$^{151,153}$Sm~\cite{simon2016}. The upbend was experimentally shown to be of dipole nature in 
$^{56}$Fe~\cite{larsen2013}. Moreover, it has been demonstrated~\cite{larsen_goriely2010} that
such a low-energy enhancement in the $\gamma$SF could significantly increase 
radiative neutron-capture rates of relevance for the $r$-process -- if found to be 
present in very neutron-rich nuclei.

In 2012, the upbend was independently confirmed in $^{95}$Mo~\cite{wiedeking2012} using a different technique.
This triggered theoretical 
investigations of the origin of this phenomenon. Within the thermal-continuum quasiparticle 
random-phase approximation (TCQRPA), the upbend was explained as due to $E1$ transitions caused by
thermal single-quasiparticle excitations in the continuum~\cite{litvinova2013}, with its strength
depending on the nuclear temperature. On the other hand, shell-model 
calculations~\cite{schwengner2013,brown2014} show a strong increase in $B(M1)$ strength 
for low-energy $M1$ transitions. At present, $^{60}$Ni is the only case where experimental data
favor a magnetic character of the upbend~\cite{voinov2010}. More experimental information is needed in order to 
determine whether the upbend is dominantly of  magnetic or electric character, or a mixture of both.

In this work, we present NLDs and $\gamma$SFs of $^{56,57}$Fe extracted from (p,p$'$$\gamma$) coincidences,
to be compared with data from other reactions using heavier projectiles, inducing higher
initial spins.
We analyze systematic errors in the normalization procedure and compare our results to available
data in the literature. For the first time, we present angular distributions of the 
upbend in $^{57}$Fe, as well as $\gamma$SFs as function of excitation energy to investigate
the so-called \textit{generalized Brink-Axel hypothesis} for $^{56,57}$Fe. This hypothesis has up to now
only been validated for the heavy nucleus $^{238}$Np~\cite{guttormsen2016}. 

This article is organized as follows. In section~\ref{sec:exp},
we give experimental details and the main steps of the Oslo-method analysis. In section~\ref{sec:nldgsf},
the NLDs and $\gamma$SFs are shown and the normalization uncertainties are discussed. 
Further, in section~\ref{sec:ang} angular distributions are presented for $^{57}$Fe, while
section~\ref{sec:brink} deals with $\gamma$SFs as function of excitation energy and implications
for the generalized Brink-Axel hypothesis. Finally, a summary and outlook are given in 
section~\ref{sec:sum}.

\section{Experimental details and data analysis}
\label{sec:exp}

The experiments were performed at the Oslo Cyclotron Laboratory (OCL). A 16-MeV proton  
beam with intensity of $\approx 0.5$ nA impinged on self-supporting targets of 
99.9\% enriched $^{56}$Fe
and 92.4\% enriched $^{57}$Fe. Both targets had mass thickness of
$\approx 2$ mg/cm$^2$. Accumulating times were $\approx 85$h and $\approx 92$h for 
$^{56,57}$Fe, respectively.

The charged ejectiles were measured
with the Silicon Ring  particle-detector system (SiRi)~\cite{siri} 
and the $\gamma$ rays with the CACTUS array~\cite{CACTUS}. 
The SiRi system consists of eight $\Delta E - E$ telescopes. Each telescope is composed of a 130-$\mu$m thick 
front detector segmented into eight strips (angular resolution of $\Delta\theta \simeq 2^{\circ}$), and
a 1550-$\mu$m thick back detector. 
In total, SiRi has 64 individual detectors and a solid-angle coverage of $\approx 6$\%. For these 
experiments, SiRi was placed in forward angles with respect to the beam direction, covering  
$40-54^{\circ}$. From the measured energy of the ejectiles and the reaction kinematics, the excitation energy
of the residual nucleus is deduced.

\begin{figure}[hb]
\begin{center}
\includegraphics[clip,width=0.65\columnwidth]{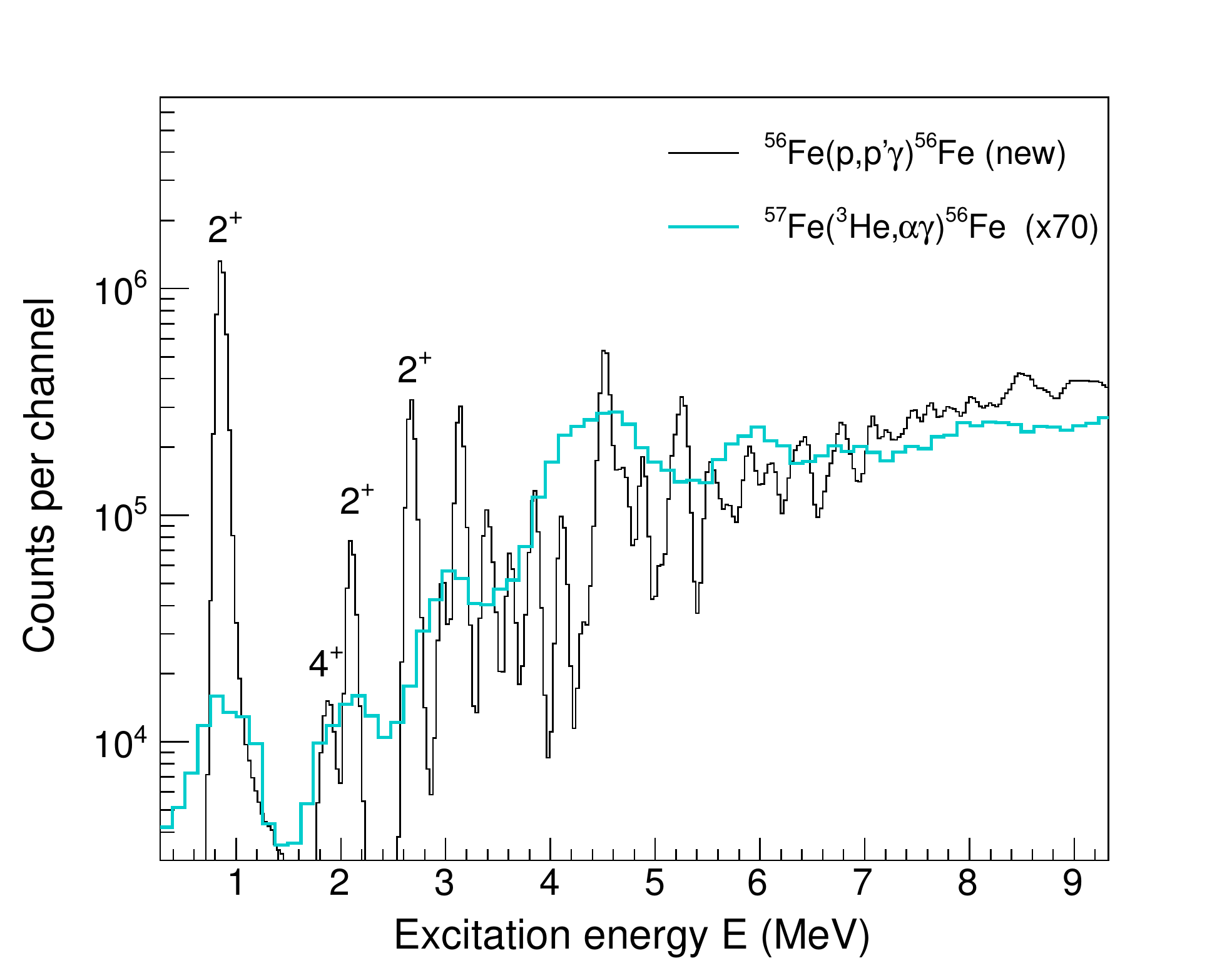}
\caption {(Color online) Proton spectra (black histogram, this work) 
and $\alpha$ spectra~\cite{voinov2004} (thick cyan line, scaled with a factor of 70) in coincidence with $\gamma$ rays 
measured with the CACTUS NaI detectors for $^{56}$Fe. Energy bins are 31 keV/channel for protons and 123 keV/channel
for $\alpha$s. The first excited levels are marked with their spin/parity.}
\label{fig:particlespectra}
\end{center}
\end{figure}

In this experiment, the CACTUS array contained 22 collimated  5 \textit{in.} $\times$ 5 \textit{in.} 
NaI(Tl) detectors and six collimated 3.5 \textit{in.} $\times$ 8 \textit{in.} LaBr$_3$(Ce) detectors
from the Milan HECTOR$^{+}$ array~\cite{nicolini2007,giaz2013}. 
The NaI detectors were placed on the CACTUS 
frame with six different angles $\theta$ with respect to the beam direction (37.4, 63.4, 79.3, 100.7, 116.6, 
and 142.6 degrees), while the LaBr$_3$ crystals covered four angles (63.4, 79.3, 100.7, and 116.6 degrees).
The $\gamma$-energy thresholds were $\approx 400$ keV and $\approx 800$ keV for the NaI and LaBr$_3$
detectors, respectively. Particle-$\gamma$ coincidences were recorded event-by-event, with the overlap of the $\Delta E$ and $E$
detectors of SiRi as mastergate for the analog electronics. 
To obtain reasonable statistical error bars, i.e. $\approx 50$\% or better on the 
extracted NLD and $\gamma$SF, about 40,000 coincidences
are needed.
In total, after background subtraction of random coincidences (about 10\% of the 
prompt time peak), 
about 65 million coincidences were obtained for the NaI detectors and about 12 million coincidences
for the LaBr$_3$ detectors with the $^{56}$Fe target. Correspondingly, for $^{57}$Fe, about
15 million and 2.1 million coincidences were recorded for the NaI and LaBr$_3$ detectors, respectively.
The time resolution of the SiRi-NaI detectors was 14.4(5) ns and for the SiRi-LaBr$_3$ detectors
6.3(3) ns.

In figure~\ref{fig:particlespectra}, the proton spectrum of SiRi in coincidence with $\gamma$ rays from the
present experiment is compared to the $\alpha$ spectrum from the previous experiment reported in 
Ref.~\cite{voinov2004}. The significant improvement in energy resolution is clear; the proton spectra 
have a full width at half maximum (FWHM) of $\approx 90$ keV compared to the $\alpha$ spectra where 
FWHM $\approx 500$ keV. 
The main reason for this improvement is the segmentation of the $\Delta E$ detectors in SiRi
compared to the old setup with a non-segmented $\Delta E$ detector. The segmentation allows for 
a much more precise determination of the scattering angle and thus the recoil energy. 
Also, using 
a proton beam instead of a $^3$He
beam gives a smaller recoil energy to the residual nucleus.
For more details, we refer to~\cite{siri}.

\begin{figure}[htb]
\begin{center}
\includegraphics[clip,width=1.09\columnwidth]{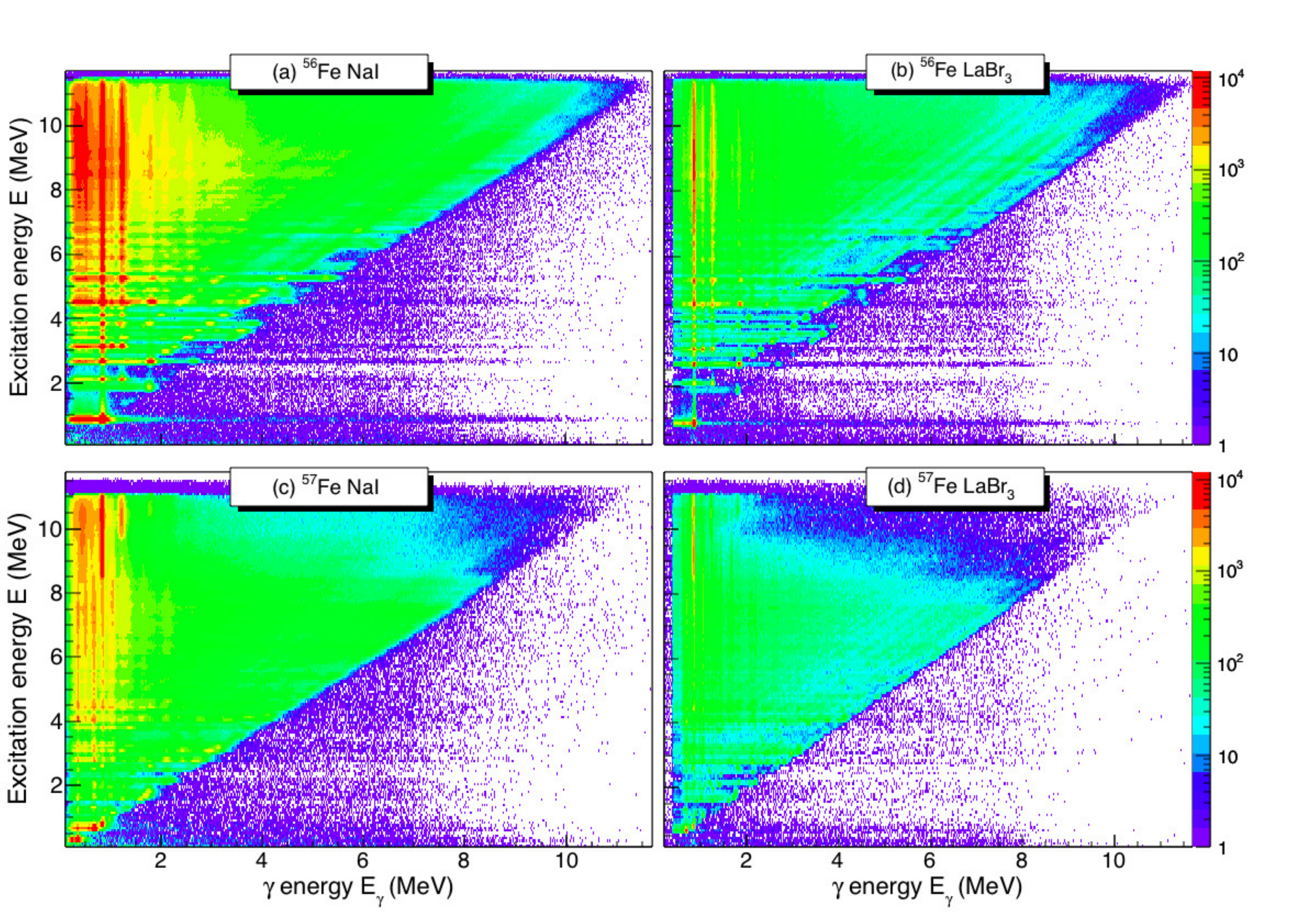}
\caption {(Color online) $\gamma$-ray energy versus excitation energy before unfolding 
for (a) $^{56}$Fe, NaI detectors; (b) $^{56}$Fe, LaBr$_{3}$ detectors; (c) $^{57}$Fe, NaI detectors;
(d) $^{57}$Fe, LaBr$_{3}$ detectors. Energy bins are 14 keV/channel.}
\label{fig:matrices}
\end{center}
\end{figure}

The proton-$\gamma$ coincidence matrices for the NaI and LaBr$_3$ detectors are displayed in 
figure~\ref{fig:matrices}. The superior energy resolution for the LaBr$_3$ spectra relative to the NaI ones
is evident, 
as well as diagonals for which the excitation energy $E$ equals the $\gamma$ energy 
$E_\gamma$ corresponding to decay to the ground state. Other diagonals are also clearly visible, for example
the direct decay to the first-excited 2$^+$ state in $^{56}$Fe. 

It is also very interesting to note the "triangles" in the $^{57}$Fe matrix where the $\gamma$ intensity suddenly drops,
see for example at $E_\gamma \approx E \approx 8.5$ MeV in figure~\ref{fig:matrices}c,d. One would naively think that the
$\gamma$ intensity would be significantly reduced as soon as the  
neutron separation energy $S_n$ is reached; however, this is well above $S_n = 7.646$ MeV. 
This feature is explained by considering the average spin $\left< J\right>$ 
populated at high excitation energies. 
From $\gamma$ transitions in coincindence with protons, we identify the decay from the $6^+$ level 
at $E = 3.39$ MeV in $^{56}$Fe as well as other levels with spins 2, 3, 4, 5~\cite{larsen2013}. 
Levels with these spins will be hindered in decaying through $s$-wave neutron emission to the $0^+$ ground state
in $^{56}$Fe. This hindrance is studied in detail for $^{95}$Mo 
and applied in a novel technique to determine spins in~\cite{wiedeking2016}.

In order to obtain the correct $\gamma$-energy distribution for each excitation-energy bin, the signals from the
NaI and LaBr$_3$ detectors must be corrected for the detector response. We applied the 
unfolding technique described in~\cite{guttormsen1996}, which is an iterative procedure using a 
strong smoothing of the Compton part of the spectrum. In order to construct response functions for
the NaI and LaBr$_3$ detectors, we used in-beam measured
transitions from $^{56}$Fe, $^{28}$Si, $^{13}$C, and $^{16}$O~\cite{crespocampo2016}. 

Moreover, we made use of a subtraction technique~\cite{guttormsen1987}
to extract the distribution of primary $\gamma$ rays, 
i.e. the first $\gamma$ rays emitted in the decay cascades, 
for each excitation-energy bin. This 
distribution contains information on the NLD and the $\gamma$SF
as deduced from Fermi's Golden Rule~\cite{dirac1927,fermi1950}:
\begin{equation}
\lambda = \frac{2\pi}{\hbar}|\left<f|H'|i\right>|^2 \rho_f,
\label{eq:fermi}
\end{equation}
where $\lambda$ is the decay rate between initial state $i$ and final state $f$, $H'$ is the transition
operator and $\rho_f$ is the density of final states. Similarly, the distribution of primary $\gamma$ rays as
function of $E$ depends on the level density at $E_f = E-E_\gamma$ and the $\gamma$-transmission coefficient
${\mathcal{T}}$
for the $\gamma$ transition with energy $E_\gamma$. The $\gamma$-transmission coefficient
is directly proportional to the $\gamma$SF. Our ansatz is~\cite{schiller2000}:
\begin{equation}
P(E_{\gamma},E) \propto  \rho (E_f) {\mathcal{T}}  (E_{\gamma}),
\label{eq:brink}
\end{equation}
where $P(E_{\gamma},E)$ is the matrix of primary $\gamma$ rays, 
representing relative intensities or branching ratios for a given 
transition energy $E_\gamma$ at a given initial excitation energy $E$. 

\begin{figure}[bt]
\begin{center}
\includegraphics[clip,width=1\columnwidth]{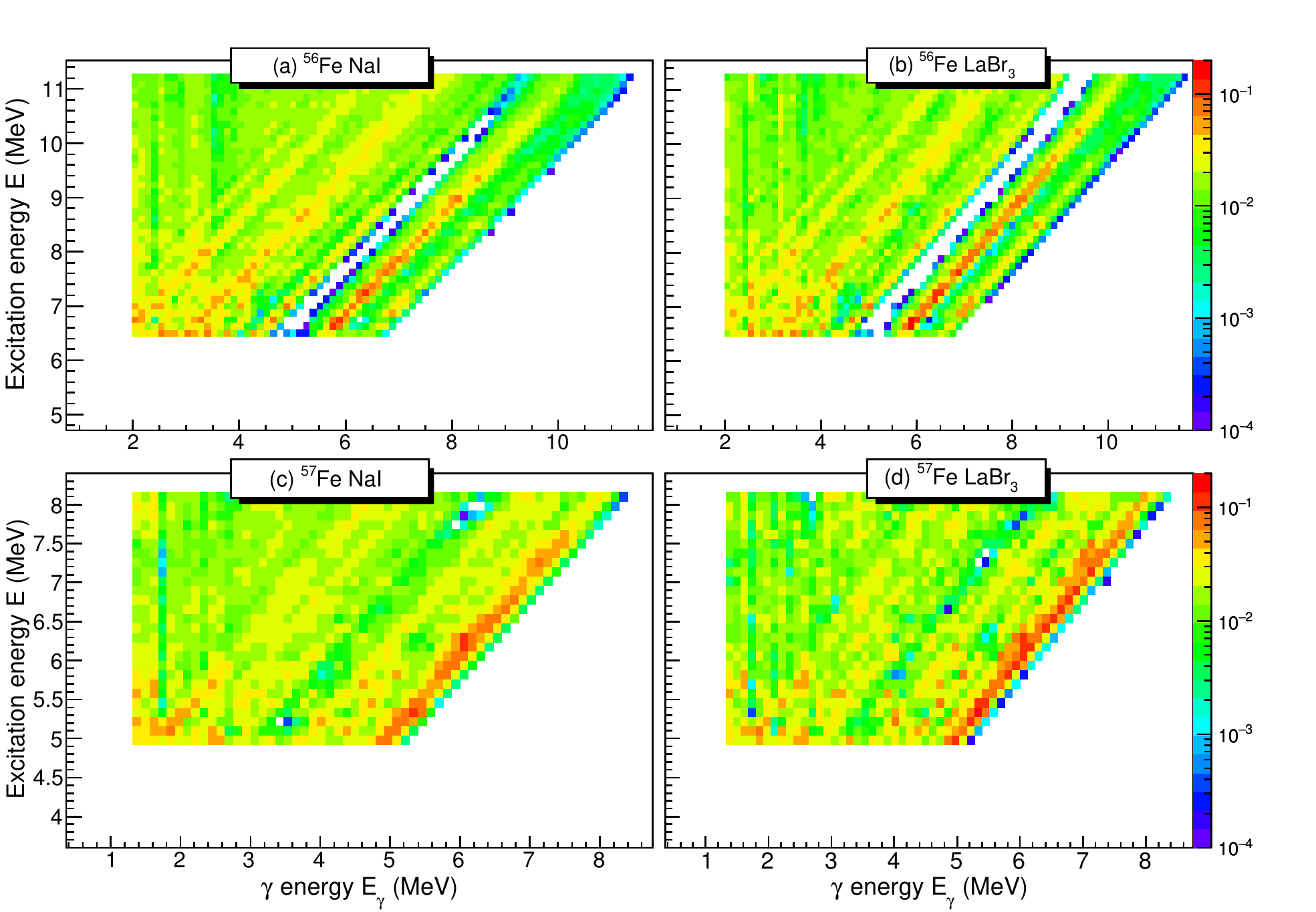}
\caption {(Color online) Distribution of primary $\gamma$ rays energy versus excitation energy 
for (a) $^{56}$Fe, NaI detectors; (b) $^{56}$Fe, LaBr$_{3}$ detectors; (c) $^{57}$Fe, NaI detectors;
(d) $^{57}$Fe, LaBr$_{3}$ detectors. Energy bins are 124 keV/channel for $^{56}$Fe and 120 keV/channel
for $^{57}$Fe. Note the different energy scales for the lower and upper panels.}
\label{fig:fg_rsg_matrices}
\end{center}
\end{figure}

The primary $\gamma$-ray matrices $P(E_{\gamma},E)$ for $^{56,57}$Fe are shown in 
figure~\ref{fig:fg_rsg_matrices}. They are normalized for each excitation-energy bin so that 
$\sum_{E_\gamma} P(E_{\gamma},E) = 1$. This means that the probability for $\gamma$ decay from a given 
bin is 1, and that the intensity of a given $\gamma$-ray energy reflects the branching ratio for
that particular transition energy.

These matrices are used as input for the extraction of the NLD and 
$\gamma$SF for the four data sets. 
The expression in equation~\ref{eq:brink} is valid for statistical decay, i.e. where the decay is independent
of the formation of the compound state~\cite{bohr-mottelson1969}. This is fulfilled at rather high
excitation energies where the initial NLD is high, typically above $\approx 2\Delta$ where 
the pair-gap parameter $\Delta\approx 12A^{-1/2}$~\cite{bohr-mottelson1969}. Note that 
${\mathcal{T}}$ is a function only of $E_\gamma$ and not $E$ or $E_f$, in accordance with the generalized 
Brink-Axel hypothesis~\cite{brink1955,axel1962}. This will be discussed in detail in section~\ref{sec:brink}.

The functional form of the NLD and $\gamma$SF is determined through a least-$\chi^2$ fit to the 
$P(E_{\gamma},E)$ matrices as described in~\cite{schiller2000}.  
The 3D landscapes
as shown in figure~\ref{fig:fg_rsg_matrices} are used in the fit. 
The sum of all primary transitions for each $E$ bin is normalized to unity.
As the 
$P(E_{\gamma},E)$ matrices contain many more data points ("pixels") than the free parameters
(the vector elements of $\rho(E_f)$ and ${\mathcal{T}}(E_\gamma)$), the solution is uniquely
determined and the fit routine converges fast, typically within 10-20 iterations. 

Some considerations need to be made before extracting the NLD and $\gamma$SF from the data. 
First, a low-energy limit for the excitation energy is applied to avoid the discrete region
at low $E$, for which the condition of a compound-nucleus decay is highly questionable. 
Further, an upper limit $E_{\mathrm{max}}$ must be given, which typically corresponds to 
$S_n$, as neutrons are not measured or discriminated in the present
experimental setup. Finally, a low-energy limit on the $\gamma$ energy, $E_{\gamma,\mathrm{low}}$,
is determined to exclude eventual higher-generation transitions not properly subtracted in the 
primary-distribution extraction, as discussed in detail in~\cite{larsen2011}. 
The chosen 
energy limits for the extraction procedure are: 
$E_{\gamma,\mathrm{low}} = 2.1$ MeV, $E_{\mathrm{min}} = 6.6$ MeV, and
$E_{\mathrm{max}} = 11.3$ MeV for $^{56}$Fe; correspondingly, $E_{\gamma,\mathrm{low}} = 1.4$ MeV, 
$E_{\mathrm{min}} = 5.0$ MeV, and $E_{\mathrm{max}} = 8.2$ MeV for $^{57}$Fe. The neutron separation
energies $S_n$ are 11.197 MeV and 7.646 MeV for $^{56,57}$Fe, respectively. The reason why we are able to 
put $E_{\mathrm{max}}$ higher than $S_n$ in the case of $^{57}$Fe, is that the first-excited
state in $^{56}$Fe is at 847 keV, allowing in principle for $E_{\mathrm{max}} = (7.65 + 0.85)$ MeV = 8.5 MeV 
as we are requiring proton-$\gamma$ coincidences. Similarly, for $^{57}$Fe, the upper limit is $\approx$ 100 keV
above $S_n$. 

\begin{figure}[bt]
\begin{center}
\includegraphics[clip,width=1.1\columnwidth]{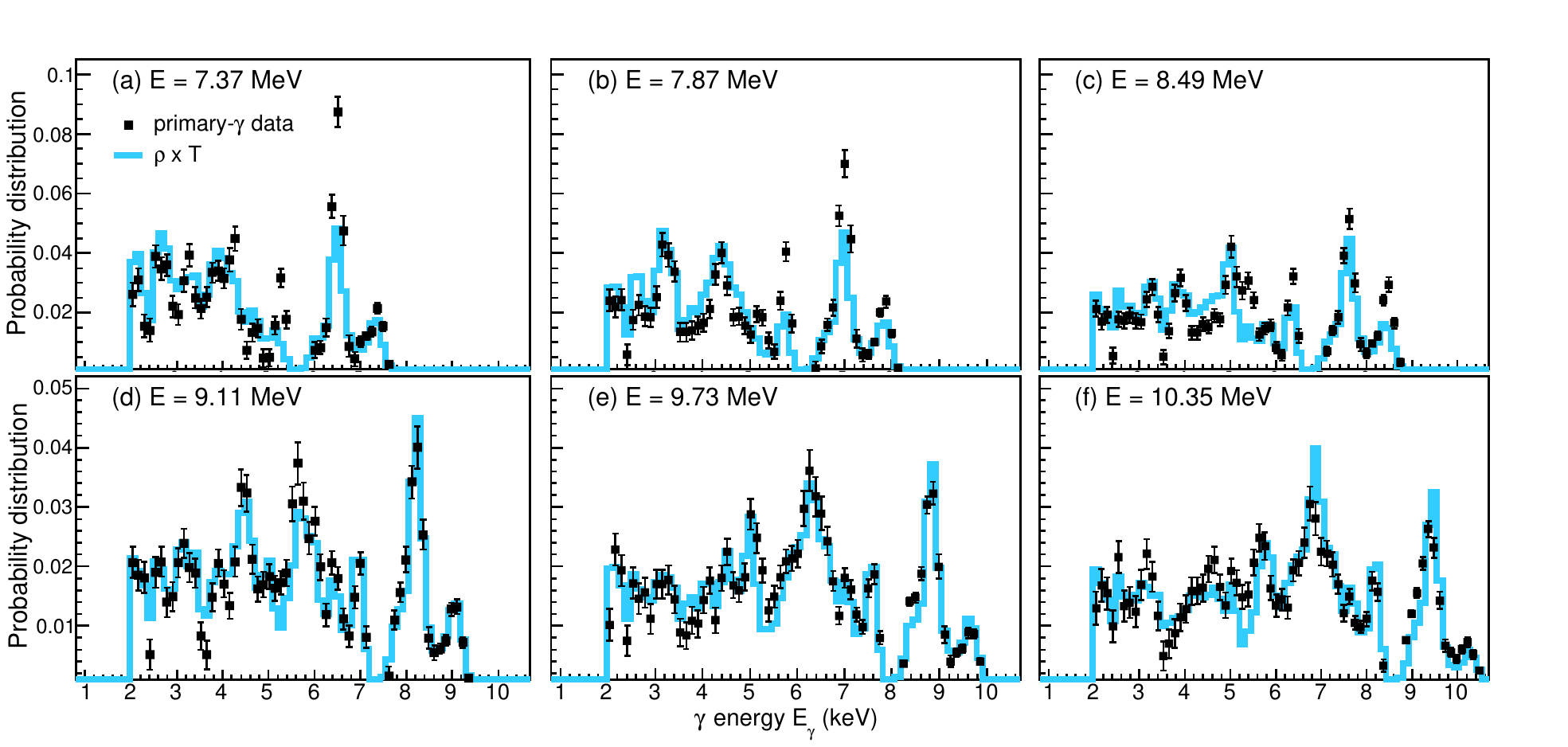}
\caption {(Color online) Comparison of experimental primary $\gamma$ spectra for $^{56}$Fe (black points, NaI detectors) with
the calculated ones (blue histogram) from the extracted $\rho$ and ${\mathcal{T}}$ functions for a set of 
initial excitation-energy bins as indicated in the panels. Energy bins are 124 keV/channel.}
\label{fig:diw_56Fe_NaI}
\end{center}
\end{figure}

To test the quality of the fit, which is based on all primary spectra included in the extraction
procedure, we take the obtained $\rho(E_f)$ and ${\mathcal{T}}(E_\gamma)$ functions and use them 
to generate primary $\gamma$ spectra to be compared with the input spectra bin by bin. 
This is shown in figure~\ref{fig:diw_56Fe_NaI}. Error bars in the primary spectra reflect
statistical uncertainties, and systematic uncertainties stemming from the unfolding procedure and
the extraction of the primary $\gamma$ rays~\cite{schiller2000}.

\begin{figure}[bt]
\begin{center}
\includegraphics[clip,width=1.1\columnwidth]{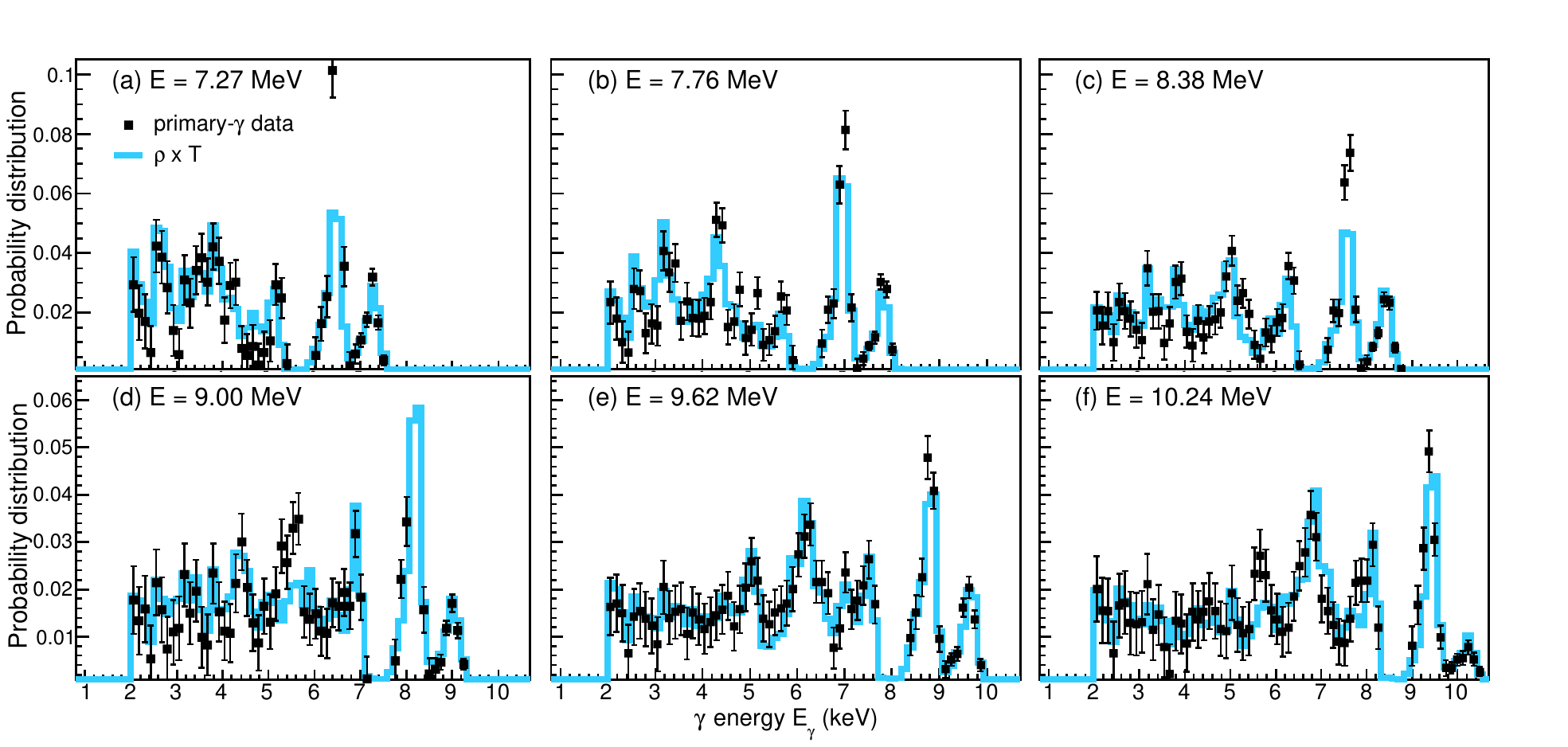}
\caption {(Color online) Same as figure~\ref{fig:diw_56Fe_NaI} for $^{56}$Fe, using data from the LaBr$_3$ detectors.}
\label{fig:diw_56Fe_LaBr3}
\end{center}
\end{figure}

\begin{figure}[bt]
\begin{center}
\includegraphics[clip,width=1.1\columnwidth]{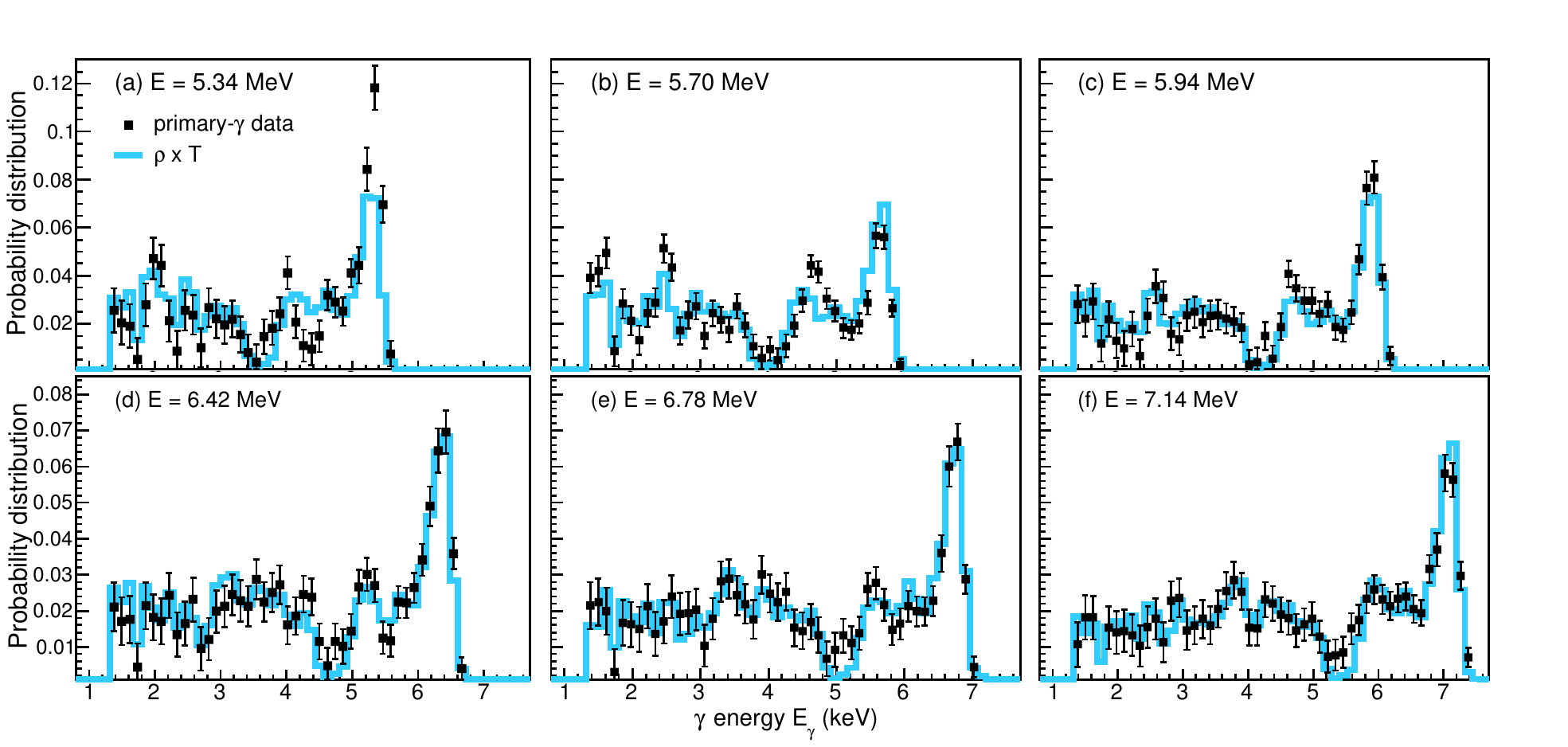}
\caption {(Color online) Same as figure~\ref{fig:diw_56Fe_NaI} for $^{57}$Fe measured with NaI detectors.
Energy bins are 120 keV/channel.}
\label{fig:diw_57Fe_NaI}
\end{center}
\end{figure}

\begin{figure}[bt]
\begin{center}
\includegraphics[clip,width=1.1\columnwidth]{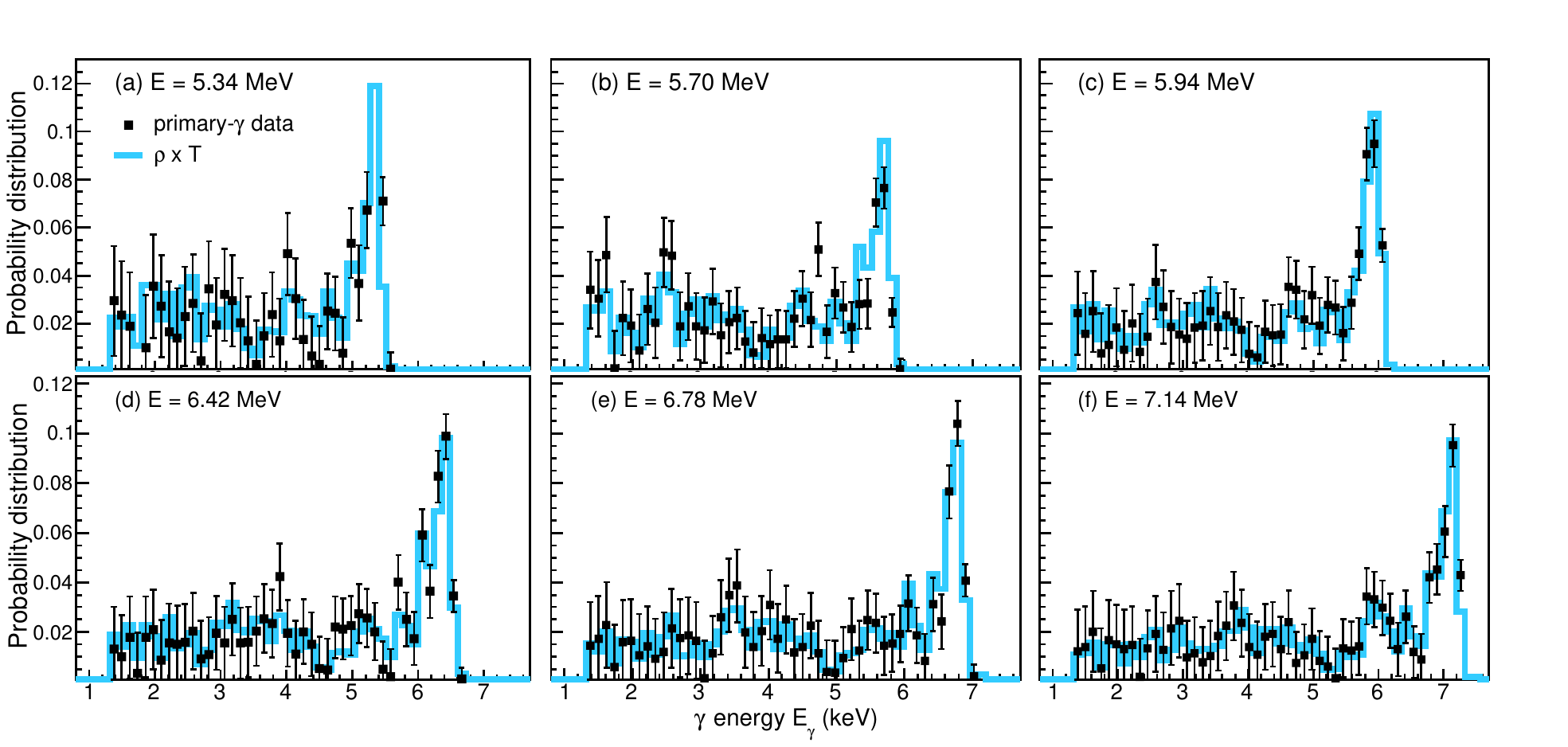}
\caption {(Color online) Same as figure~\ref{fig:diw_56Fe_NaI} for $^{57}$Fe using data from the LaBr$_3$ detectors.}
\label{fig:diw_57Fe_LaBr3}
\end{center}
\end{figure}

As can be seen from figures~\ref{fig:diw_56Fe_NaI}--\ref{fig:diw_57Fe_LaBr3}, 
the overall agreement between the data and the calculated primary spectra is very good. It should be noted
that Porter-Thomas fluctuations~\cite{porter-thomas} of the decay strengths are not taken into account. 
These fluctuations are expected to be large when the final level density $\rho_f$ is low. This is 
clearly visible \textit{e.g.} in the decay to the first-excited level in $^{56}$Fe, see figure~\ref{fig:diw_56Fe_NaI}a
and the peak at $E_\gamma \approx 6.5$ MeV, where data points are several standard deviations off the calculated 
$\rho \times \mathcal{T}$. Here, there is only
one final level and the relative decay strength is seen to fluctuate strongly for different initial 
excitation energies.

\section{Level density and $\gamma$ strength}
\label{sec:nldgsf}

\subsection{Normalization}
As only the functional form of the NLD and $\gamma$SF can be deduced from the primary $\gamma$
spectra, the slope and absolute normalization must be determined from auxiliary data. It is 
shown in~\cite{schiller2000} that any solution $\tilde{\rho_f}$ and $\tilde{{\mathcal{T}}}$ will give an
equally good $\chi^2$ fit to the primary-$\gamma$ data through the transformations 
\begin{eqnarray}
\rho(E-E_\gamma)&=&\mathcal{A}\exp[\alpha(E-E_\gamma)]\,\tilde{\rho}(E-E_\gamma),
\label{eq:array1}\\
{\mathcal{T}}(E_\gamma)&=&\mathcal{B}\exp(\alpha E_\gamma)\tilde{{\mathcal{T}}} (E_\gamma),
\label{eq:array2}
\end{eqnarray}
where the parameters $\mathcal{A}$, $\mathcal{B}$, are the absolute normalization
of the NLD and the $\gamma$-transmission coefficient, respectively, and $\alpha$ is the common slope parameter. 

For the NLD, the parameters $\mathcal{A}$ and $\alpha$ are found by fitting our data
to known levels from the literature~\cite{NNDC} at low excitation energy and to 
neutron-resonance spacing data from~\cite{RIPL3} at $S_n$. 
The discrete levels are binned with the same bin width as our experimental data.
For $^{56}$Fe, there is no 
information from neutron-resonance experiments as $^{55}$Fe is unstable. For this case,
we have estimated the NLD at $S_n$ from systematics in the following way: 
\begin{itemize}
\item[(\textit{i})]{To estimate the lower-limit NLD, 
we calculate the total level density from the $s$-wave neutron resonance spacing
$D_0$ for Fe isotopes where this value is available from~\cite{RIPL3} according to the 
expression 
\begin{equation}
\rho(S_n) = \frac{2\sigma^2}{D_0} \cdot \frac{1}{(J_t+1)\exp\left[-(J_t+1)^2/2\sigma^2\right] + J_t\exp\left[-J_t^2/2\sigma^2\right]},
\label{eq:rhoSn}
\end{equation}
assuming equally many positive- and negative-parity states. Here, 
 $J_t$ is the ground-state spin of the target nucleus in the neutron-resonance experiment and $\sigma$
is the spin cutoff parameter. We make use of the phenomenological spin cutoff parameter suggested 
in~\cite{egidy2009}:
\begin{equation}
\sigma^2(E) = 0.391 A^{0.675}(E - 0.5Pa^{\prime})^{0.312}. 
\label{eq:spincut}
\end{equation}
Here, $A$ is the mass number and $Pa^{\prime}$ is the deuteron pairing energy 
as defined in~\cite{egidy2009}. This approach gives a low value for the spin cutoff parameter and thus a low 
limit for the level density. Further, we calculate $\rho(S_n)$ from the global systematics~\cite{egidy2009}
directly. By taking the $\chi^2$ fit of the semi-experimental $\rho(S_n)$ with the values from systematics 
in the same fashion as done for $^{89}$Y in~\cite{larsen2016}, one obtains an estimate
for the $^{56}$Fe $\rho_{\mathrm{low}}(S_n)$. All parameters are given in table~\ref{table1}. This normalization
is referred to as \textit{norm-1} in the following.}
\item[(\textit{ii})]{To estimate the upper-limit NLD, we apply the same procedure as in (\textit{i}) but
with the spin cutoff parameter given by the rigid-body moment of inertia approach as parameterized in~\cite{egidy2005}:
\begin{equation}
\sigma^2(E) = 0.0146A^{5/3}\frac{1+\sqrt{1+4a(E-E_1)}}{2a}.
\label{eq:spincut05}
\end{equation}
Here, $a$ is the level-density parameter and $E_1$ is the excitation-energy backshift determined from global systematics of
~\cite{egidy2005}. All parameters are given in table~\ref{table2}. We refer to this normalization
as \textit{norm-2}.}
\end{itemize}

For $^{57}$Fe, we use the $D_0$ value given in~\cite{RIPL3} and estimate $\rho(S_n)$ using equation~\ref{eq:rhoSn}, 
again with spin cutoff parameters both from~\cite{egidy2009} and~\cite{egidy2005}. 
Consistent with the approach for $^{56}$Fe, the lower limit is obtained with the spin cutoff parameter
in equation~\ref{eq:spincut}, and the upper limit with the one in equation~\ref{eq:spincut05}, also including the uncertainties 
in $D_0$. 
All parameters are listed in 
table~\ref{table1} and~\ref{table2} in~\ref{appendixA}.

\begin{figure}[bt]
\begin{center}
\includegraphics[clip,width=1.1\columnwidth]{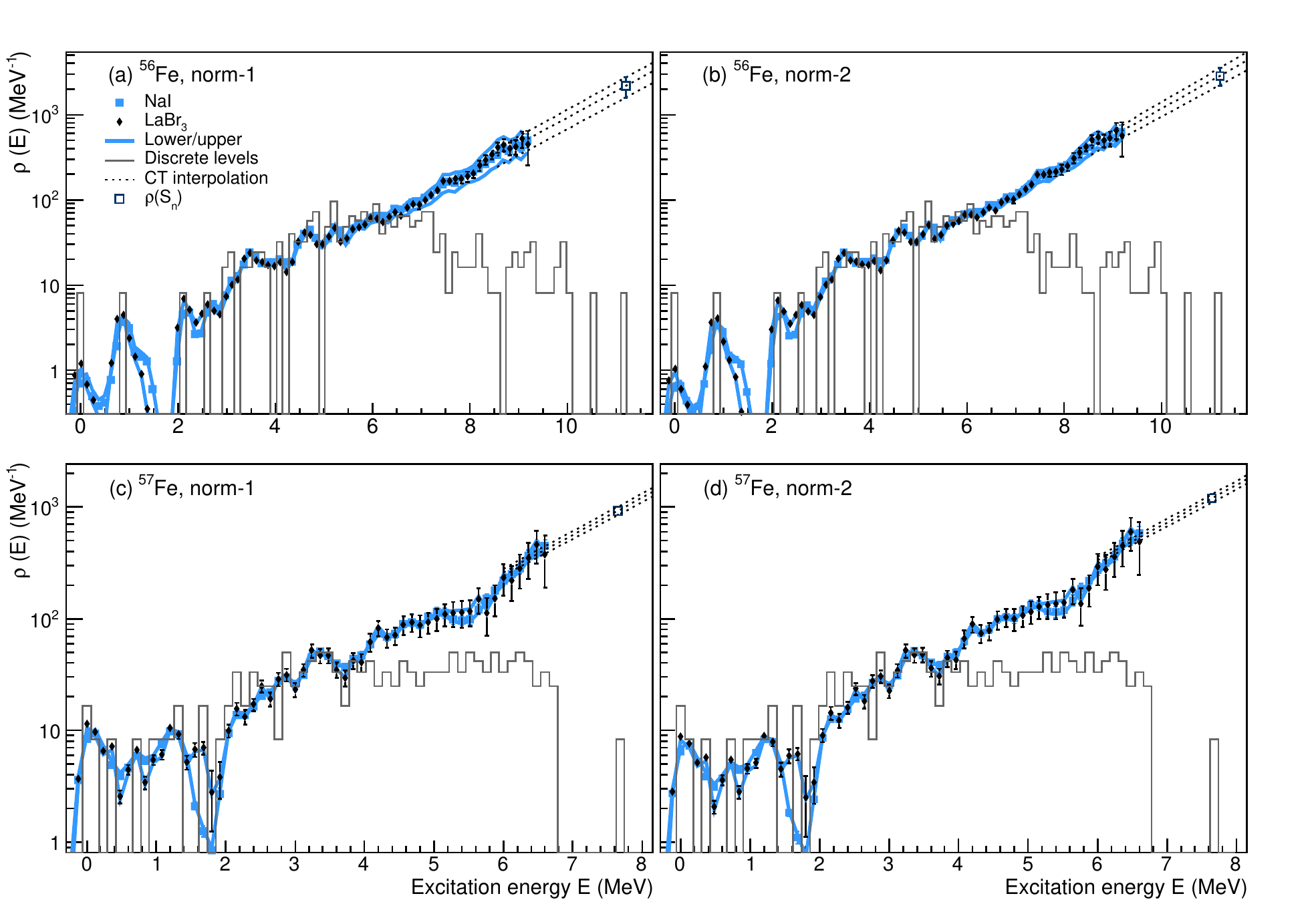}
\caption {(Color online) Normalized level densities for (a) $^{56}$Fe, norm-1, (b) $^{56}$Fe, norm-2, 
(c) $^{57}$Fe, norm-1, and (d) $^{57}$Fe, norm-2.}
\label{fig:nld}
\end{center}
\end{figure}

As our data reach up to $E_{\mathrm{max}} - E_{\gamma,\mathrm{low}}$, we must interpolate between the estimated
$\rho(S_n)$ and our upper data points. This is done using the constant-temperature formula of 
Ericson~\cite{ericson1959,ericson1960}:
\begin{equation}
\rho_{CT}(E) = \frac{1}{T}\exp{\frac{E-E_0}{T}}.
\label{eq:rhoCT}
\end{equation}
The applied parameters $T$ and $E_0$ are given in table~\ref{tab:ctvalues} for the various normalization options,
giving the best fit to our data in the regions $E=8.2-9.2$ MeV and $E=6.2-6.6$ MeV for $^{56,57}$Fe, respectively.
The normalized level densities are shown in figure~\ref{fig:nld}.

With the normalized NLDs at hand, and assuming equal parity~\cite{voinov2001},
we normalize the $\gamma$-ray transmission coefficient $\mathcal{T}$ to the average, total
radiative width $\left<\Gamma_{\gamma0}\right>$ taken from~\cite{RIPL3} (see table~\ref{table1}) according to~\cite{voinov2001}
\begin{eqnarray}
\langle \Gamma_{\gamma0}(S_n,J_t\pm 1/2,&\pi_t)\rangle =
\frac{\mathcal{B}}{4\pi\rho(S_n,J_t\pm 1/2,\pi_t)}\int_{E_{\gamma}=0}^{S_n}\mathrm{d}E_{\gamma}\mathcal{T}(E_{\gamma}) \nonumber \\ 
&\times \rho(S_n-E_{\gamma}) \sum_{J= -1}^{1} g(S_{n}-E_{\gamma},J_{t}\pm 1/2+J),
\label{eq:width}
\end{eqnarray}
where $J_t$ and $\pi_t$ are the spin and parity of the target nucleus in the $(n,\gamma)$ reaction and
$\rho(S_n-E_{\gamma})$ is the experimental NLD. Note that 
the experimental transmission coefficient in principle includes all types of electromagnetic transitions:
$\mathcal{T}_{E1}+\mathcal{T}_{M1}+\mathcal{T}_{E2}+ ...$; however, dipole transitions are found to
be dominant for decay in the quasicontinuum (e.g.,~\cite{larsen2013,kopecky_uhl_1990}).
The sum in equation~\ref{eq:width} runs over all final states with spins 
$J_t \pm 1/2 + J$, where $J= -1,0,1$ from considering the spins 
reached after one primary dipole transition with energy $E_\gamma$ (see also equation~3.1 in~\cite{kopecky_uhl_1990}).  
Note that the factor
$1/\rho(S_n,J_t\pm 1/2,\pi_t)$ equals the neutron resonance spacing $D_0$. From the normalized transmission coefficient,
the $\gamma$SF is determined by
\begin{equation}
f(E_\gamma) = \frac{\mathcal{T}(E_{\gamma})}{2\pi E_\gamma^3}.
\label{eq:dipolestrength}
\end{equation}

\begin{figure}[bt]
\begin{center}
\includegraphics[clip,width=1.1\columnwidth]{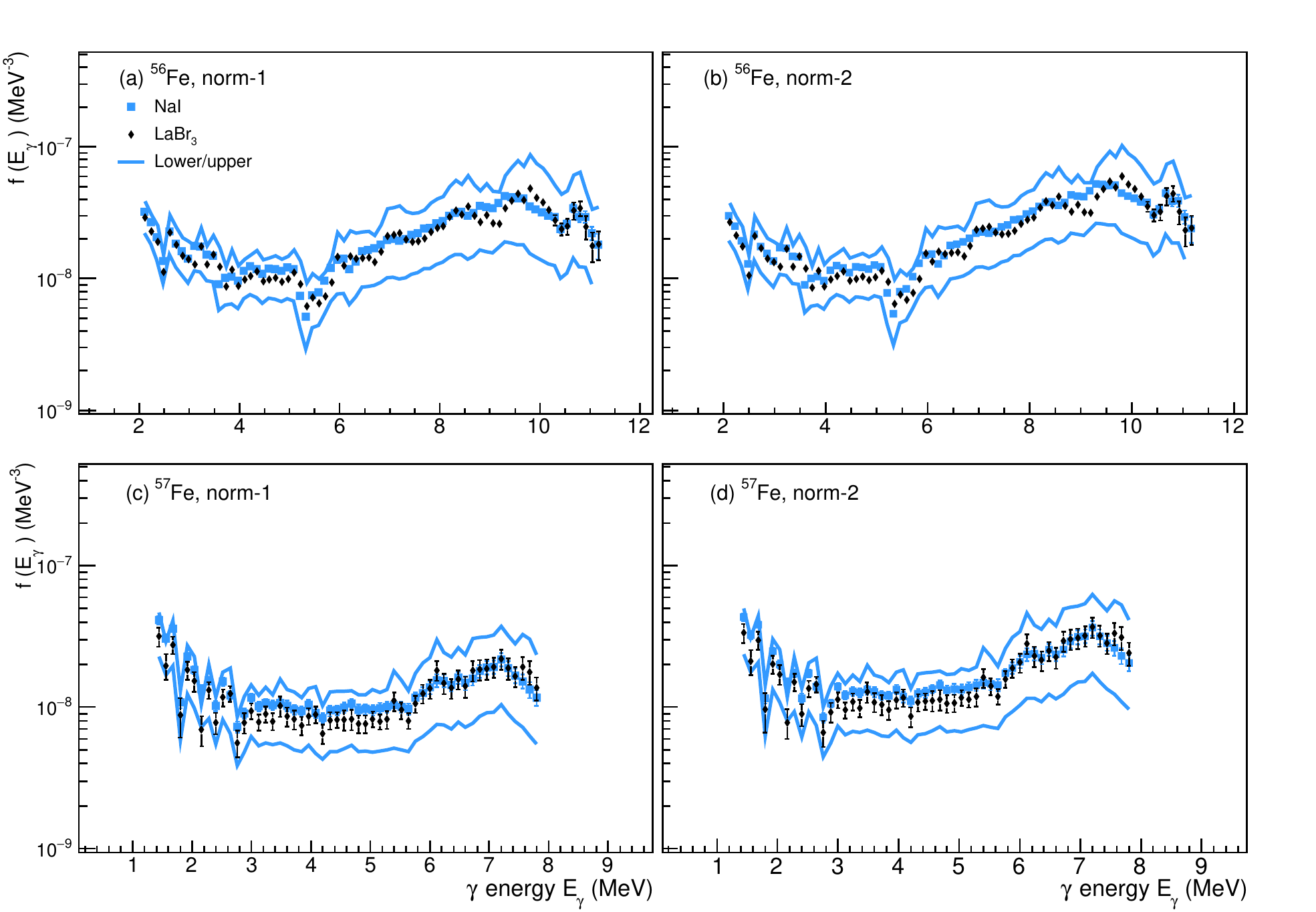}
\caption {(Color online) Normalized $\gamma$SFs for (a) $^{56}$Fe, norm-1, (b) $^{56}$Fe, norm-2, 
(c) $^{57}$Fe, norm-1, and (d) $^{57}$Fe, norm-2.}
\label{fig:gsf}
\end{center}
\end{figure}

Again, $^{56}$Fe lacks neutron resonance data and we have therefore estimated $\left<\Gamma_{\gamma0}\right>$ from a linear 
fit to the values of the other Fe isotopes taken from~\cite{RIPL3}, see table~\ref{table1}. The normalized $\gamma$SFs
for the different normalization options for the level densities are shown in figure~\ref{fig:gsf}. The error band
includes uncertainties in $D_0$,  spin cutoff parameters, and $\left<\Gamma_{\gamma0}\right>$.
We see that the $\gamma$SFs have a distinct U-like shape, independent on the choice of normalization. 
There is a characteristic increase in strength at low transition energies, which is very similar in shape
and magnitude to recent predictions from large-scale shell-model calculations~\cite{brown2014}.

At the highest $\gamma$-ray energies, we observe a drop in strength, which could be due to the reaction 
populating spins at high excitation energies that on average are higher than the (close-to) ground-state spin(s),
and/or a small overlap with the wave functions for the initial and final levels.
In particular, for $^{56}$Fe, only $1^-$ and $1^+$ levels contribute to the dipole strength to the ground state. 
For lower transition energies, a broad range of levels is available as the final level density is much higher. 
One should therefore note that 
the upper data points ($E_\gamma > 9.5$ and 7.2 MeV for $^{56,57}$Fe, respectively) do not represent a general, averaged
$\gamma$SF in the quasicontinuum. The rather peculiar behavior of these data points indicate a possible (strong) dependence 
on the initial and final level(s), as well as significant Porter-Thomas fluctuations. 
This will be further investigated and discussed in section~\ref{sec:brink}. 

\subsection{Comparison with other data}
There exist data on the NLDs of $^{56,57}$Fe from previous experiments at the OCL~\cite{voinov2004}, using
the $^3$He-induced reactions $^{57}$Fe($^{3}$He,$\alpha\gamma$)$^{56}$Fe and 
$^{57}$Fe($^{3}$He,$^{3}$He$^\prime\gamma$)$^{57}$Fe. 
Moreover, level densities have also been inferred from particle-evaporation spectra
of the reactions $^{55}$Mn(d,n)$^{56}$Fe~\cite{voinov2006}, $^{59}$Co(p,$\alpha$)$^{56}$Fe~\cite{vonach1966},
$^{58}$Fe($^{3}$He,$\alpha$)$^{57}$Fe~\cite{voinov2007}, and $^{60}$Ni(n,$\alpha$)$^{57}$Fe~\cite{fischer1984}.
Reactions involving heavier projectiles and/or ejectiles, as well as detection angles in
backward direction where the compound-reaction mechanism is dominant, populate
higher initial spins than for the ($p,p^\prime$) reaction in forward angles. 
This would help nailing down eventual reaction dependencies on the final results.

\begin{figure}[bt]
\begin{center}
\includegraphics[clip,width=1.1\columnwidth]{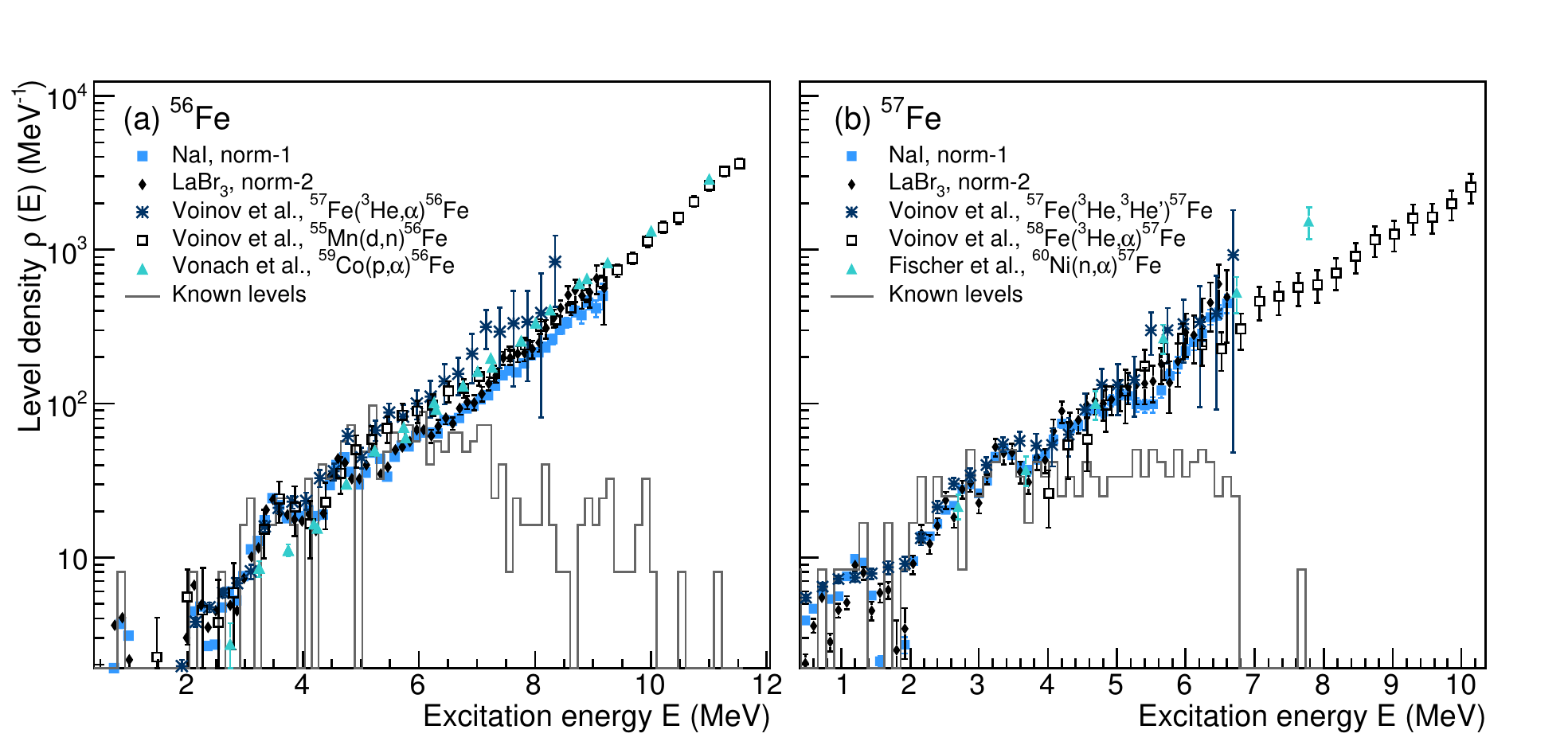}
\caption {(Color online) Comparison of NLDs from different reactions for (a) $^{56}$Fe and (b) $^{57}$Fe. 
Previous data taken from ~\cite{voinov2004,voinov2006,vonach1966,voinov2007,fischer1984}.}
\label{fig:nldcomp}
\end{center}
\end{figure}

Figure~\ref{fig:nldcomp} shows the comparison of the present data and previous results on the NLDs. 
We find that the overall agreement is very good, although there are some differences betweeen the data sets. 
For $^{56}$Fe, 
we see that the particle-evaporation data give a higher NLD between $E \approx 5-7.5$ MeV,
which is interpreted as due to the higher spins reached in these experiments compared to the 
proton inelastic scattering.
The absolute normalization of our data is rather uncertain due to the lack 
of neutron-resonance data as discussed before; 
however, there is a significant boost in the number of levels at $E \approx 6$ MeV for all data sets relative to the 
known, discrete levels. 
For $^{57}$Fe, a similar increase is taking place at $E \approx 4$ MeV. 
This could be caused by two factors: a quenching of pair correlations due to breaking of 
nucleon Cooper pairs, and sufficient energy to cross the $f_{7/2}$ shell gap with more than one
particle (neutron or proton) into the $p_{3/2},f_{5/2},p_{1/2}$ orbitals.

We note that there is a significant deviation between the data of 
\cite{voinov2007} and \cite{fischer1984} above $S_n$ for $^{57}$Fe. 
It would be highly desirable to perform new experiments
in this energy region to clarify whether this is due to different spins populated, the particle
transmission coefficients used in the analyses or issues with their absolute normalization to the 
discrete levels.
\begin{figure}[bt]
\begin{center}
\includegraphics[clip,width=1.1\columnwidth]{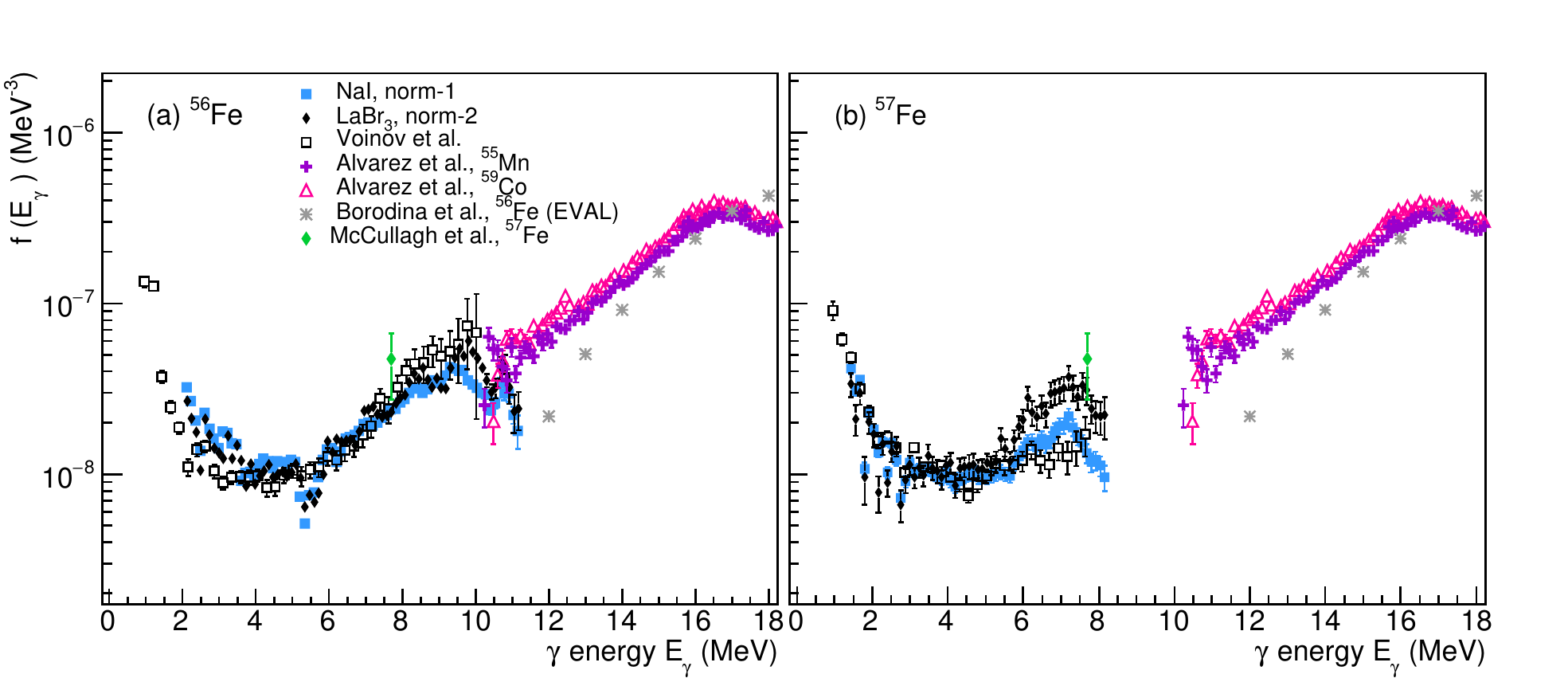}
\caption {(Color online) Comparison of $\gamma$SFs from different reactions for (a) $^{56}$Fe and (b) $^{57}$Fe. 
Photonuclear data are taken from~\cite{alvarez1979}, and the evaluated $^{56}$Fe($\gamma,n$) cross section
from~\cite{borodina2000}. Also the $^{57}$Fe $E1+M1$ strength at $E_\gamma \approx 7.7$ MeV from resonant
($\gamma,n$) neutron time-of-flight measurements are shown~\cite{jackson1979,mccullagh1981,RIPL2}.
For $^{56}$Fe, the present work provides the $\gamma$SF for 
$2.1 \leq E_\gamma \leq 11.3$ MeV, while data from~\cite{voinov2004} cover $1.0 \leq E_\gamma \leq 10.3$ MeV.
Correspondingly, for $^{57}$Fe, the present work covers the range $1.4 \leq E_\gamma \leq 8.2$ MeV, 
and data from~\cite{voinov2004}
$1.0 \leq E_\gamma \leq 7.6$ MeV. The photonuclear data from~\cite{alvarez1979} are for $E_\gamma > 10.2$ MeV.}
\label{fig:gsfcomp}
\end{center}
\end{figure}

For the $\gamma$SF, there are to our knowledge no photonuclear data available for $^{56,57}$Fe. 
We have therefore compared our data to photoneutron ($\gamma,n$) cross sections of $^{55}$Mn and 
$^{59}$Co~\cite{alvarez1979}, and also an evaluation of the $^{56}$Fe($\gamma,n$) 
cross section~\cite{borodina2000}.
The photoneutron cross section $\sigma_{\gamma n}$ 
maps out the shape of the 
Giant Dipole Resonance (GDR)~\cite{dietrich_berman}, and 
is converted to $\gamma$ strength by the
relation~\cite{bartholomew1972}
\begin{equation}
f(E_\gamma) = \frac{1}{3\pi^2 \hbar^2 c^2}\frac{\sigma_{(\gamma,n)}(E_\gamma)}{ E_\gamma}.
\end{equation}
Moreover, data from threshold ($\gamma,n$) neutron-time-of-flight experiments on 
$^{57}$Fe~\cite{jackson1979,mccullagh1981,RIPL2} provide an estimate for the $E1$ and $M1$ strength function
at $E_\gamma \approx 7.7$ MeV; the sum of these are also compared to our data.
The result is shown in figure~\ref{fig:gsfcomp}, where we show only our normalizations for norm-1 and norm-2
for clarity.
In general, we observe a very good agreement with the previous $^3$He-induced data below $S_n$.
We note that for $^{56}$Fe, the slope of the $\gamma$SF of~\cite{voinov2004} is somewhat steeper, leading to 
an overall lower strength for $E_\gamma<4$ MeV and a higher strength above $E_\gamma \approx 7$ MeV.
This is likely due to a different (steeper) NLD normaliza{}tion as seen in figure~\ref{fig:nldcomp}a.
For $^{57}$Fe, a significant difference is only seen for $E_\gamma$ above $\approx 6$ MeV,
where the $^3$He-induced data undershoot the present results. The reason for this discrepancy is 
not clear; it could be related partly to different normalizations of the NLD, and/or strong transitions
to the ground band for the $(p,p^\prime)$ reaction that are less pronounced for the $^{3}$He inelastic scattering.

Finally, we note that our data show a natural continuation of the GDR tail
towards lower $\gamma$ energies. We stress again that the strength of the highest transition energies
close to the neutron threshold is very likely to be strongly dependent on the initial and final state(s).
Hence this is not quasicontinuum decay, but is still a real effect due to nuclear structure and spin selection
rules. For another similar case, namely $^{89}$Y($p,p^\prime$)~\cite{larsen2016}, there exist also
inelastic photon scattering data~\cite{benouaret2009}, which display the same pattern as our data
at high transition energies. This suppressed strength could be relevant also for e.g. ($n,\gamma$) 
cross-section calculations, if the neutron-capture reaction populates the same low initial spins. 
Thus one would expect a suppression of primary transitions to e.g. the ground state in the neutron-capture
reaction as well.

According to the principle of detailed balance~\cite{blatt_weisskopf}, one expects
that the photoabsorption ("upward") strength equals the decay ("downward") strength. However, as discussed by 
Bartholomew \textit{et al.}~\cite{bartholomew1972}, the principle of detailed balance is only strictly fulfilled 
within the extreme statistical model. Hence, one would assume that this is only valid at high excitation
energies where the NLD is high and the wave functions are strongly mixed. Moreover, the Brink hypothesis
has been used~\cite{brink1955} for calculating radiative widths at $S_n$, again assuming that the GDR
tail extrapolated from photonuclear data to low $\gamma$ energies could appropriately describe 
the decay process. Despite rather large uncertainties, it is fair to say that the 
present experimental data from different reactions agree reasonably well and provide a quite consistent 
picture on the general shape of the $\gamma$SF. 

\subsection{Comparison with theory}
\begin{figure}[!ht]
\begin{center}
\includegraphics[clip,width=1.\columnwidth]{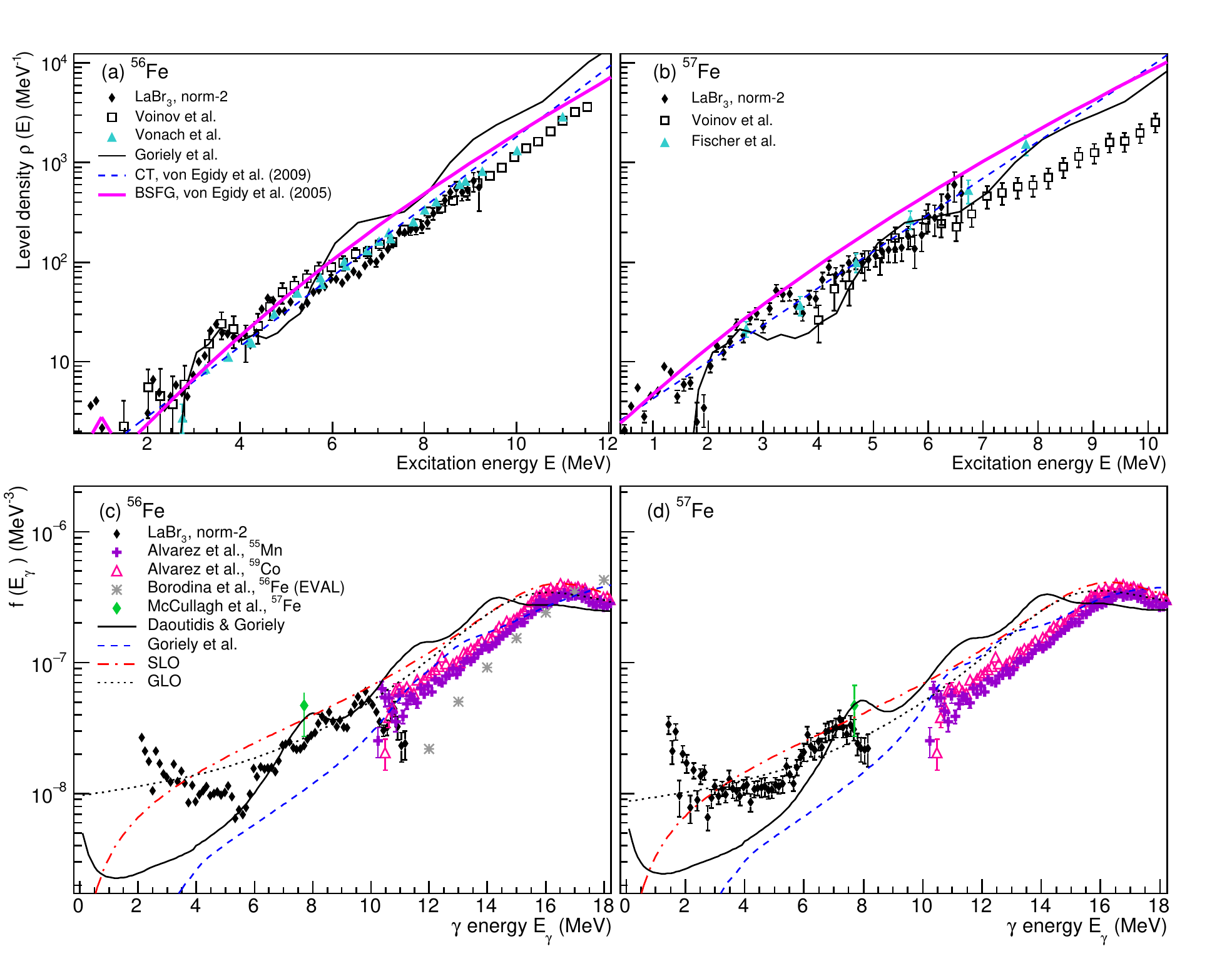}
\caption {(Color online) Comparison of data and theoretical calculations for 
(a) $^{56}$Fe NLD, (b) $^{57}$Fe NLD, (c) $^{56}$Fe $\gamma$SF and (d) $^{57}$Fe $\gamma$SF. 
Microscopic level densities within the Hartree-Fock-Bogoliubov plus combinatorial approach 
are from~\cite{goriely2008}. Parameters for the CT model and the back-shifted Fermi-gas (BSFG) model 
are from~\cite{egidy2009} and~\cite{egidy2005}, respectively. For the $\gamma$SF models,
GDR parameters for the standard Lorentzian (SLO)~\cite{brink1955} and the
Generalized Lorentzian (GLO)~\cite{kopecky_uhl_1990} are from~\cite{RIPL2}. The microscopic 
$\gamma$SF calculations are from~\cite{daoutidis_goriely2012} (taken at $T=1.4$ MeV) and \cite{goriely2004}.}
\label{fig:theorycomp}
\end{center}
\end{figure}

Although there are many phenomenological models and some more microscopic calculations available,
they typically deviate considerably both in shape and magnitude. In Fig.~\ref{fig:theorycomp} a
selection of frequently used models are compared to the data. 
Note that we have used the global parameterization for the NLD models of~\cite{egidy2009}
and~\cite{egidy2005} to test their predicitve power. 
For the NLD, none of the models 
reproduce the data over the full energy range. Clearly, only the microscopic approach is 
able to grasp some of the structures seen in the experimental results. Apparently, all the NLD
models overshoot the data at high excitation energy. This could have severe consequences for
e.g. calculations of reaction cross sections.

Also for the $\gamma$SF models, the situation is rather confusing. Again, there is no
model that can capture all features for the full $\gamma$-energy range. The smooth, phenomenological 
$E1$ models are quite appropriate close to the GDR, but are missing the upbend at low transitions
energies. Even for the GLO model, which actually overshoots our $^{56}$Fe data 
at $E_\gamma \approx 4-7$ MeV, underestimates the strength by a factor of $\approx 3$ 
at $E_\gamma = 2.1$ MeV. This might not be a surprise, considering the possible 
$M1$ nature of the low-energy enhancement~\cite{schwengner2013,brown2014}, or an enhancement
due to strong $E1$ continuum single-particle transitions~\cite{litvinova2013}, none of 
which are incorporated in the phenomenological models. 
Also, the microscopic models are undershooting 
the low-energy data as well--the $^{56}$Fe data show a factor of $\approx 30$ more strength than
the quasi-particle random-phase approximation (QRPA) $E1$ strength~\cite{goriely2004}.
However, the microscopic approaches do show
structural features rather similar to the data between $7-10$ MeV. Clearly, the situation
is at present far from satisfactory, and more theoretical work is required to understand
in depth both NLDs and $\gamma$SFs, preferably within the same theoretical framework. \\

\section{Angular distributions, $^{57}$Fe}
\label{sec:ang}
In~\cite{larsen2013}, it was shown that the low-energy upbend in $^{56}$Fe is dominated by dipole transitions. 
Here, we apply the same type of analysis for the so-far unexplored $^{57}$Fe upbend. 

We use the various angles for which the NaI detectors are placed and extract angular distributions 
by sorting the data into $(E_\gamma,E)$ matrices according to $\theta$ of the NaI detectors relative 
to the beam direction. As the LaBr$_3$ detectors were placed at only four angles,
and  had a rather high $E_\gamma$ threshold, these were not used for this analysis. 
From the intensities as a function of angle, 
we can fit angular-distribution functions of the form~\cite{litherland1961,mateosian1974} 
\begin{equation}
W(\theta) = A_0 + A_2 P_2(\cos \theta) + A_4 P_4 (\cos \theta),
\label{eq:legendre}
\end{equation}
where $P_k(\cos \theta)$ is a Legendre polynomial of degree $k$.

The normalized angular-distribution coefficients are given by $a_k = Q_k \alpha_k A_k/A_0$, where $Q_k\approx 1$ is the 
geometrical attenuation coefficient due to the finite size of the $\gamma$ detectors, and $\alpha_k$ is the attenuation 
due to partial alignment of the nuclei relative to the beam direction. We estimate uncertainties in the intensities according to
$\mathbf{\sigma_{\mathrm{tot}} = \sqrt{\sigma^2_{\mathrm{stat}} + \sigma^2_{\mathrm{syst}}}}$. 
The statistical errors are given by 
$\sqrt{N}$ where $N$ is the number of counts, and the systematic errors are deduced from the relative change in $N$
for each symmetric pair of angles (37.4$^\circ$,142.6$^\circ$), 
(63.4$^\circ$,116.6$^\circ$), and (79.3$^\circ$,100.7$^\circ$). 
The statistical errors are typically $\approx 4$\% or smaller,
and the systematic errors are thus the dominant source of uncertainty.
More details about the angular distributions are given in~\ref{AppendixB}.

In figure~\ref{fig:angular_trans} we
show the angular distributions of known transitions in $^{57}$Fe, and how they compare with the 
theoretical $a_k^{\mathrm{max}}$ values. All numbers are given in table~\ref{tab:coeff}.
The comparison with the experimentally extracted $a_2$ coefficients and the theoretical
maximum values for the known transitions shown in figure~\ref{fig:angular_trans}a,b, 
indicates an attenuation $\alpha_k \approx 0.6-0.75$.
\begin{figure}[bt]
\begin{center}
\includegraphics[clip,width=0.9\columnwidth]{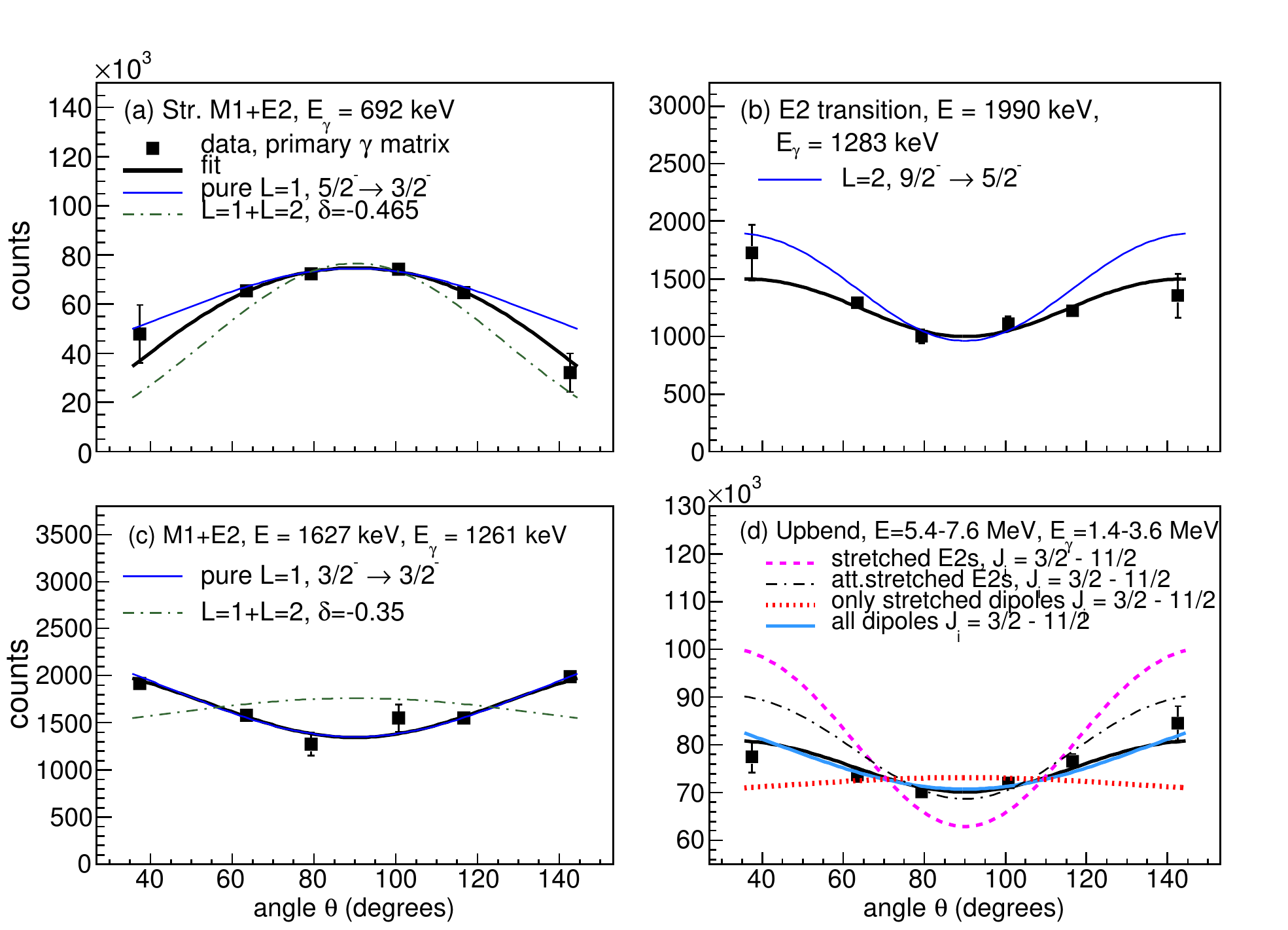}
\caption {(Color online) Angular distributions for (a)--(c) single transitions and (d) the upbend region from primary transitions in $^{57}$Fe.}
\label{fig:angular_trans}
\end{center}
\end{figure}
\begin{table*}[tb]\footnotesize
\caption{Angular-distribution coefficients of transitions measured in the present experiment.
The theoretical $a_k^{\mathrm{max}}$ coefficients for complete alignment are taken from Ref.~\cite{mateosian1974}.} 
\begin{tabular}{cccccrrrr}
\hline
\hline
$E$    & $E_\gamma$  & $I_i \rightarrow I_f$     & $XL$    & $\delta$ & $a_2^{\mathrm{max}}$ & $a_2$       & $a_4^{\mathrm{max}}$ & $a_4$       \\
 (keV) & (keV)       &                           &         &          &                      &             &                      &             \\
\hline
   706 &  692        & $5/2^- \rightarrow 3/2^-$ & $M1+E2$ & $-0.465$ & $-1.068$             & $-0.80(20)$ & 0.12                 & $-0.11(11)$  \\
  1627 & 1261        & $3/2^- \rightarrow 3/2^-$ & $M1+E2$ & $-0.35$  & $-0.127$             & $0.35(5)$   & $0.00$               & $-0.05(14)$  \\
  1990 & 1283        & $9/2^- \rightarrow 5/2^-$ & $E2$    & $-$      & $0.476$              & $0.28(20)$  & $-0.29$              & $-0.21(16)$  \\

\hline
\end{tabular}
\\
\label{tab:coeff}
\end{table*}

The behavior of the $E_\gamma = 1261$ keV 
non-stretched\footnote{Transitions are called stretched for a maximum change in the 
angular momentum of the nuclear states, and non-stretched if the change is less than 
the maximum allowed for the given multipolarity.} $M1+E2$ transition is somewhat puzzling, as~\cite{ensdf}
gives a rather large mixing parameter of $-0.35$ (see figure~\ref{fig:angular_trans}c). 
The shape of our data indicates a stronger contribution
from the non-stretched $M1$ part, although we do have a large uncertainty in the $a_4$ parameter.
Nevertheless, assuming a pure $M1$ transition, one finds $a_2^{\mathrm{max}} = 0.400$, which is close to the 
experimental value of $0.35(5)$. 
We note that all $a_4$ parameters are consistent with 0 within their error bars.

For the upbend, we have fitted equation~\ref{eq:legendre} to the primary spectra for the range
$E=5.4-7.6$ MeV and $E_\gamma = 1.4-3.6$ MeV with $a_2$ and $a_4$ as free parameters, obtaining
$a_2 = 0.11(6)$ and $a_4 = -0.06(6)$ (see figure~\ref{fig:angular_trans}d). 
The uncertainty in $a_4$ is very large, but its value is small,
indicating that contributions from stretched $E2$ transitions are not dominant. Moreover, 
we have made a fit of the data to the sum of Legendre polynomials for $J_i = 3/2 - 11/2$, 
with a weighting coefficient for the stretched and the non-stretched part. Here, we obtain 
65(12) and 35(6)\% for the non-stretched and the stretched transitions, respectively.
Note that possible contributions from other spins and $E2$ transitions
could modify these numbers, which should only be taken as a qualitative guidance.

However, when we fit only the sum of the stretched $E2$ Legendre polynomials for $J_i = 3/2 - 11/2$,
we find a significantly worse agreement, see figure~\ref{fig:angular_trans}d. This is
true also for the fit including the maximum experimental attenuation of $\approx 0.60$, 
Fitting a sum of stretched and non-stretched dipoles and stretched $E2$ transitions
yields a fit similar to that of only the attenuated $E2$s. Also a fit of only
stretched dipole transitions is clearly not reproducing the data.
The best fit (more than a factor of 5 better $\chi^2$) is obtained with a sum of 
stretched and non-stretched dipole transitions, although small $E2$ contributions
e.g. from $M1+E2$ mixing cannot be ruled out.

\begin{figure}[bt]
\begin{center}
\includegraphics[clip,width=1.1\columnwidth]{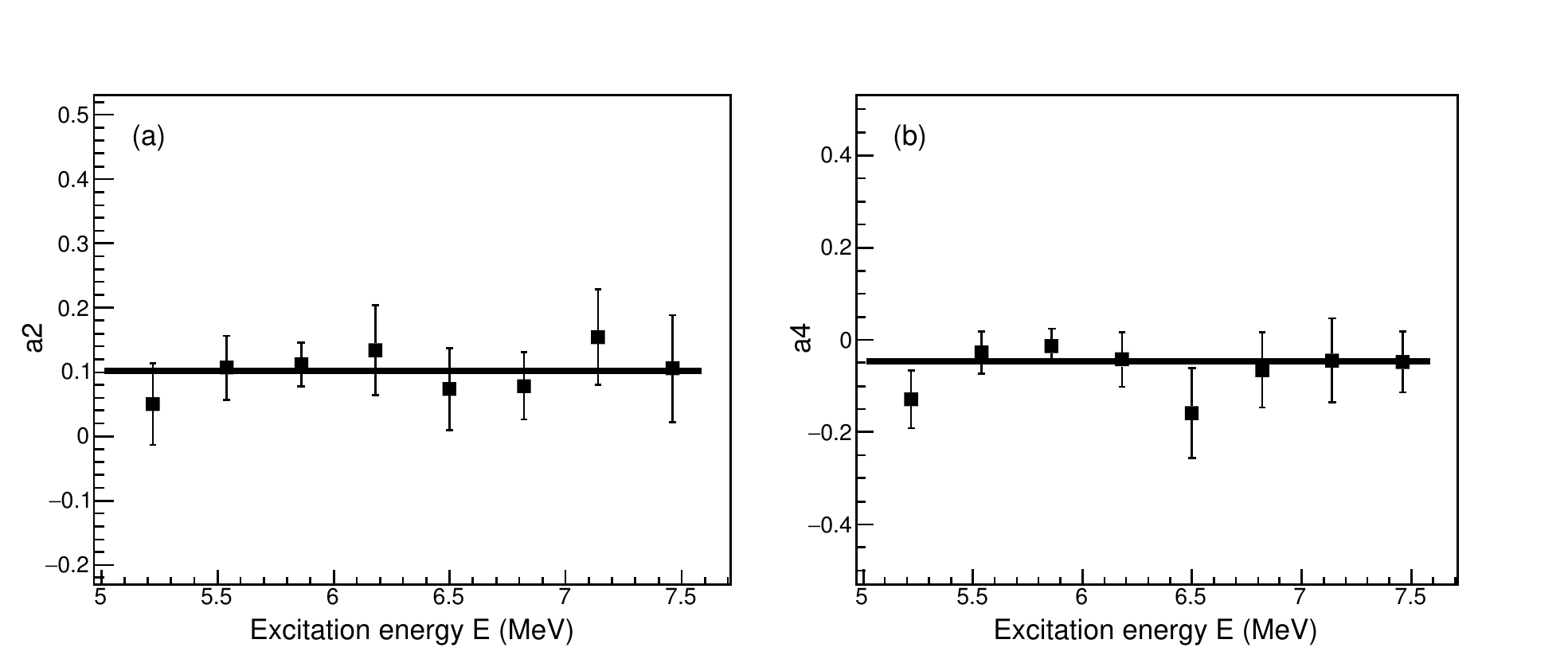}
\caption {Extracted $a_2$ and $a_4$ coefficients from independent fits of 300-keV excitation-energy cuts in the 
$^{57}$Fe primary matrices for the six CACTUS angles.}
\label{fig:angcoeff}
\end{center}
\end{figure}

To study the angular-distribution coefficients for the upbend in $^{57}$Fe in more detail, we  make individual fits
of equation~\ref{eq:legendre} to eight 300-keV wide excitation-energy cuts in 
the primary $\gamma$-ray matrix in the range  
$E=5.4-7.6$ MeV, $E_\gamma = 1.4-3.6$ MeV. 
The resulting $a_2$ and $a_4$ coefficients are shown in figure~\ref{fig:angcoeff}.
We obtain $\mathbf{a_2 = 0.10(2)}$ and $\mathbf{a_4 = -0.05(2)}$, in excellent agreement with 
the simultaneous fit to the whole region as shown in 
figure~\ref{fig:angular_trans}d. 

The same trend was found in theoretical $\left< B(M1)\right>$ values from shell-model calculations 
of $^{57}$Fe~\cite{brown2014}, where non-stretched $M1$ transitions contributed most to 
the low-energy enhancement. Also, stretched $M1$ transitions dominated both experimentally~\cite{larsen2013}
and theoretically~\cite{brown2014} in the case of $^{56}$Fe, 
bringing together a consistent picture, at least qualitatively.
Hence, we conclude that the upbend structure in $^{57}$Fe is also caused by 
dipole transitions, but for this case the non-stretched transitions seem to dominate. 

\section{Generalized Brink-Axel hypothesis: $\gamma$SF as function of excitation energy}
\label{sec:brink}
As the LaBr$_{3}$ detectors have excellent energy resolution and efficiency for high-energy $\gamma$ rays, we make use of the 
technique described in~\cite{guttormsen2005,guttormsen2016,guttormsen2011} to extract the $\gamma$SF as function of excitation energy. 

We start with the primary $\gamma$-ray matrix $P(E_\gamma,E)$ obtained in section~\ref{sec:exp}. We will now 
make the assumption that the NLD is the one determined in section~\ref{sec:nldgsf}, but 
the transmission coefficient $\mathcal{T}$ is now allowed to be dependent on both
excitation energy and $\gamma$-ray energy$, \mathcal{T}(E_\gamma,E)$. As
$\rho(E_f)$ is known, we can in principle determine $\mathcal{T}(E_\gamma,E)$ for each excitation-energy bin just by 
dividing the primary $\gamma$ matrix with the NLD: $\mathcal{T}(E_\gamma,E) \sim P(E_\gamma,E)/\rho(E_f)$, using our ansatz in 
equation~\ref{eq:brink}. Specifically, we have 
\begin{equation}
{\rho(E - E_{\gamma})}{\cal T} (E_{\gamma},E) = N(E) P(E_{\gamma},E),
\end{equation}
where $N(E)$ is a normalization factor in units MeV$^{-1}$, depending only on the initial excitation energy. 
\begin{figure}[bt]
\begin{center}
\includegraphics[clip,width=0.8\columnwidth]{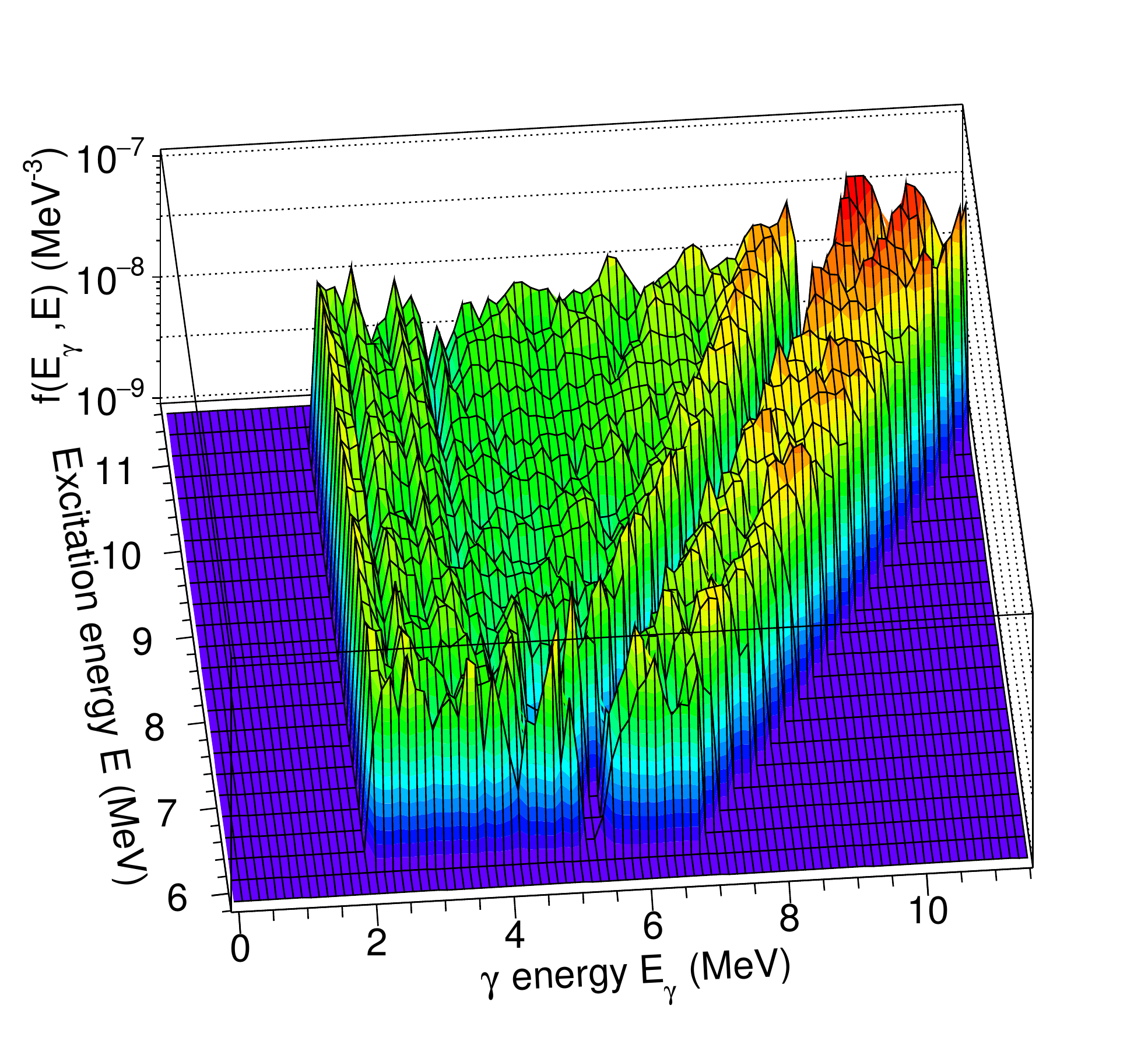}
\caption {(Color online) Extracted $\gamma$SFs as function of initial excitation energy for $^{56}$Fe. 
Bins are 248 keV/channel for  $E$ and 124 keV/channel for $E_\gamma$.}
\label{fig:radex56Fe}
\end{center}
\end{figure}

\begin{figure}[bt]
\begin{center}
\includegraphics[clip,width=0.8\columnwidth]{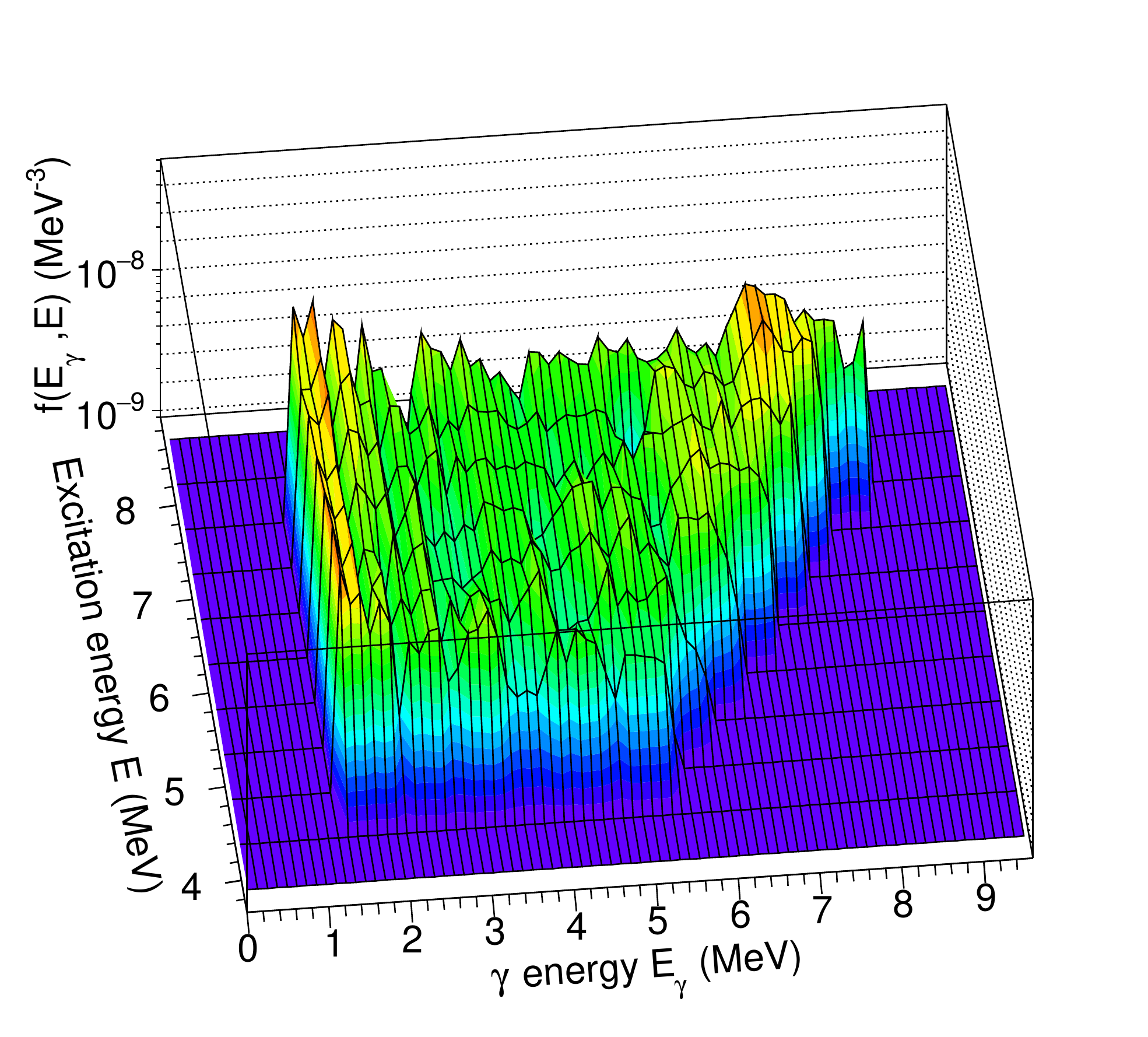}
\caption {(Color online) Extracted $\gamma$SFs as function of initial excitation energy for $^{57}$Fe. 
Bins are 480 keV/channel for $E$ and 120 keV/channel for $E_\gamma$.}
\label{fig:radex57Fe}
\end{center}
\end{figure}

Now, this game can be played in two ways:
\begin{itemize}
\item[(a)]{We investigate $\mathcal{T}$ as function of \textit{initial} excitation energy through the relation
\begin{equation}
{\cal T} (E_{\gamma},E) = N(E)
\frac{P(E_{\gamma},E)}{\rho(E - E_{\gamma})}.
\label{eq:Ei}
\end{equation}
We determine $N(E)$ by
\begin{equation}
N(E)=\frac{\int_0^{E}  {\cal T}(E_{\gamma}) \rho(E-E_{\gamma})\, {\mathrm{d}} E_{\gamma }
}{\int_0^{E} \, P(E_{\gamma},E)\, {\mathrm{d}} E_{\gamma}}.
\label{eq:nei}
\end{equation}
Note that ${\cal T}(E_{\gamma})$ is the normalized transmission coefficient from section~\ref{sec:nldgsf}. However, it will not
influence the \textit{shape} of the extracted ${\cal T} (E_{\gamma},E)$ as it acts as a constant after integrating over all
$E_\gamma$. Hence, it only serves to provide an approximate absolute normalization of ${\cal T} (E_{\gamma},E)$.
}
\item[(b)]{We can also find $\mathcal{T}$ as function of \textit{final} excitation energy by
\begin{equation}
{\cal T} (E_{\gamma},E_f) = N(E_{\gamma}+E_f)
\frac{P(E_{\gamma},E_f + E_{\gamma})}{\rho(E_f)}, 
\label{eq:Ef}
\end{equation}
where we keep in mind that $E_f + E_{\gamma} = E$. 
Again, we assume that ${\cal T}(E_{\gamma})$ gives a good estimate of the absolute value and 
we can approximate the normalization for a given final excitation energy $E_f$ and for a 
specific $E_\gamma$ fulfilling $E = E_f + E_\gamma$ by
\begin{equation}
N(E_{\gamma} + E_f)=\frac{\int_0^{E_f + E_{\gamma}}  {\cal T}(E_{\gamma}^\prime) \rho(E_f)\, {\mathrm{d}} E_{\gamma}^\prime 
}{\int_0^{E_f + E_{\gamma}} \, P(E_{\gamma}^\prime,E_f + E_{\gamma}^\prime)\, {\mathrm{d}} E_{\gamma}^\prime}.
\label{eq:nef}
\end{equation}
}
\end{itemize}
The $\gamma$SF as function of excitation energy is then easily calculated from the transmission coefficient by use
of equation~\ref{eq:dipolestrength}. The results are shown for $^{56,57}$Fe in figures~\ref{fig:radex56Fe} and
\ref{fig:radex57Fe}, respectively.

We observe that the decay strength to the ground state increases as function of both $E$ and $E_\gamma$, 
which is fully consistent with the $\gamma$SF determined previously in section~\ref{sec:nldgsf} and the expected
influence of the tail from the GDR. Moreover, we find that the $\gamma$SF varies with 
initial excitation energy, but that the general shape is preserved: there is always an upbend at low
$E_\gamma$ and a rather flat distribution of strength in the middle $E_\gamma$ region, before it
again increases for high $E_\gamma$. 

To investigate the fluctuations, following~\cite{guttormsen2011}, 
we compare the average $\gamma$SF for all initial excitation energies
with the $\gamma$SF obtained for a specific excitation-energy bin. We find that the fluctuations relative to the 
average $\gamma$SF can be large, more than 100\% for some $\gamma$-ray energies and $E$. Also,
the fluctuations are in some cases significantly larger than the error bars. Therefore, 
it seems that although the overall shape of the $\gamma$SF is indeed preserved in agreement with 
the generalized Brink-Axel hypothesis, the $\gamma$SF for a specific transition energy and excitation 
energy could have a large deviation, in particular when the excitation-energy bin is narrow
and containing rather few levels. 
\begin{figure}[bt]
\begin{center}
\includegraphics[clip,width=0.7\columnwidth]{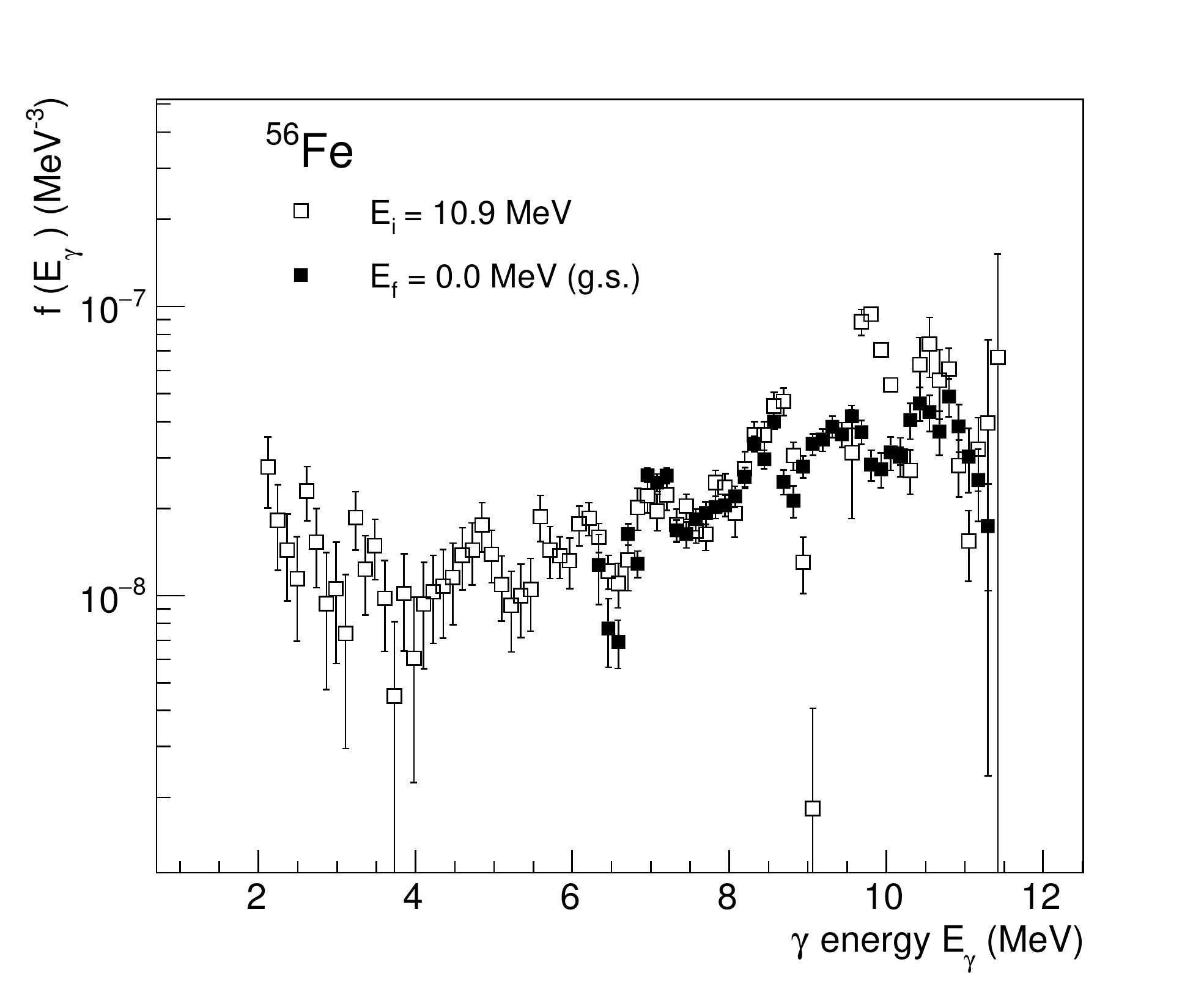}
\caption {Extracted $\gamma$SF for the $^{56}$Fe ground state as $E_f$ (black points) and for $E = 10.9$ MeV.}
\label{fig:rsf_final_56Fe}
\end{center}
\end{figure}

\begin{figure}[bt]
\begin{center}
\includegraphics[clip,width=0.7\columnwidth]{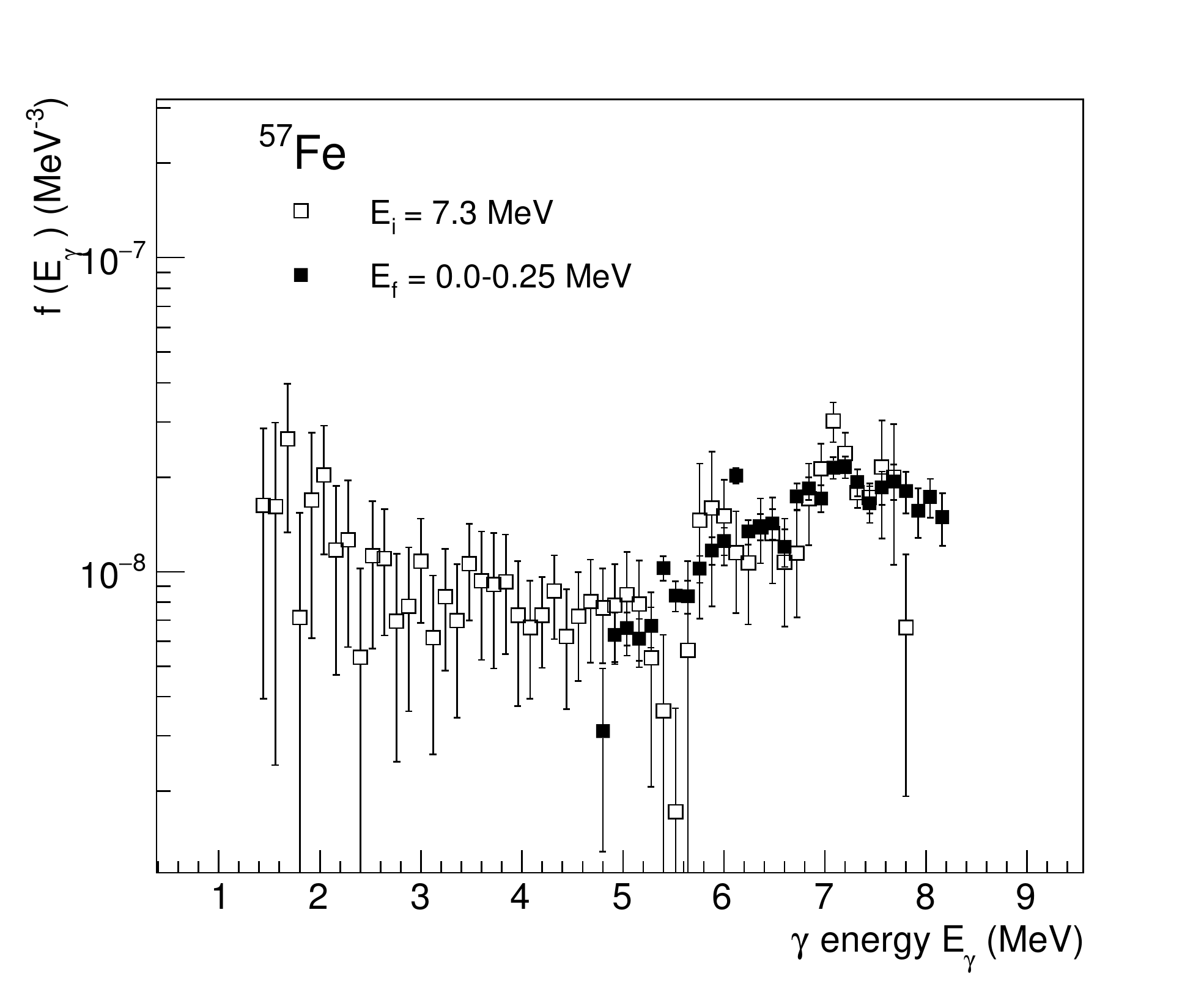}
\caption {Extracted $\gamma$SF for the $^{57}$Fe ground-state band, $E_f = 0.0-0.25$ MeV (black points) and for $E_i = 7.3$ MeV.}
\label{fig:rsf_final_57Fe}
\end{center}
\end{figure}
Finally, we also investigate the $\gamma$SF for a specific final excitation energy. We have chosen the ground state in $^{56}$Fe
and the ground-state band (1/2$^-$, 3/2$^-$, 5/2$^-$) in $^{57}$Fe. The $\gamma$SF for this $E_f$ is then compared to a typical
$\gamma$SF at a high initial $E$, see figures~\ref{fig:rsf_final_56Fe} and~\ref{fig:rsf_final_57Fe}. Again, we observe that the
general trend is preserved, although significant deviations are present, for example for the $^{56}$Fe strength at
$E_\gamma \approx 9.7$ MeV. This is interpreted to be caused by Porter-Thomas fluctuations, which are expected to be large
when the final and/or the initial NLD is low~\cite{guttormsen2016}.

\section{Summary and outlook}
\label{sec:sum}
We have presented data on $^{56,57}$Fe from ($p,p^\prime\gamma)$ reactions using
NaI and LaBr$_3$ crystals simultaneously. 
We confirm the upbend in these isotopes, which represents an increase in strength 
of a factor $\approx3-30$ relative to commonly used models for the $E1$ strength at low transition energies. 
Moreover, external data involving heavier projectiles and/or ejectiles are typically within the 
error bars. The lower NLD in $^{56}$Fe for $E \approx 5-7.5$ MeV compared to the external data
is attributed to the lower spins reached in the ($p,p^\prime$) reaction.
 
We have shown angular distributions of the upbend for $^{57}$Fe for the first time.
Our results indicate a mix of stretched and non-stretched dipoles contributing to the upbend, 
in agreement with recent shell-model calculations. Moreover, we
have investigated the excitation-energy dependence of the $\gamma$SF. The data show that the general
trends are preserved
in accordance with the Brink-Axel hypothesis. 
However, we also encounter large fluctuations, which seem to be due to strong Porter-Thomas fluctuations
caused by the low level density in these light nuclei.

Currently, the CACTUS array is in the process of being replaced by OSCAR (Oslo SCintillator ARray),
for which all the NaI detectors are replaced with 3.5 \textit{in.} $\times$ 8 \textit{in.} LaBr$_3$(Ce) detectors,
and a new frame and target chamber are being built as well. This new array will open up a wealth of new 
opportunities, such as discriminating against neutrons above $S_n$ to extract the $\gamma$SF at even higher 
energies, gating on discrete transitions
to study the feeding pattern and thus spin dependencies of the NLD and $\gamma$SF, and many more. 
We expect to be able to 
study the upbend and Porter-Thomas fluctuations in much more detail with this new equipment.

Comparing our data with frequently used NLD and $\gamma$SF models clearly shows the 
need for better theoretical predictions. After all, here we present data on stable nuclei;
the predictive power for unstable, highly exotic nuclei involved in e.g. the nucleosynthesis
is definitely unsatisfactory. To improve the situation, more data on both stable and unstable 
nuclei are required to help testing and constraining available calculations, as well
spurring new theoretical approaches and methods in nuclear physics.

\ack
The authors wish to thank J.C.~M{\"{u}}ller, E.A.~Olsen, A.~Semchenkov and J.~C.~Wikne at the 
Oslo Cyclotron Laboratory for providing excellent experimental conditions. 
This work was financed by the Research Council of Norway (NFR), project grant no. 205528, and 
through ERC-STG-2014 under grant ggreement no. 637686. 
S.~S. and G.~M.~T. acknowledge financial support
by the NFR under project grant no. 210007 and 222287, respectively.
A.~V.~V. acknowledges funding from the Department of Energy, grant no. DE-NA0002905.
M.~W. acknowledges support by the National Research Foundation of South Africa under grant no. 92789.

\appendix
\section{Data tables for normalizations}
\label{appendixA}
\begin{table*}[!h]\footnotesize
\begin{center}
\caption{Neutron resonance parameters $D_0$ and $\left< \Gamma_{\gamma 0}\right>$ from~\cite{RIPL3}, 
and spin cutoff parameters from global systematics of~\cite{egidy2009};
$A_f$ is the final nucleus following neutron capture, $J_t$ is the ground-state spin of the target nucleus, $S_n$ is the neutron-separation energy, 
$D_0$ is the $s$-wave level spacing~\cite{RIPL3},
$\sigma$ is the spin-cutoff parameter from equation~(\ref{eq:spincut}), $Pa^{\prime}$ is the deuteron shift as defined in~\cite{egidy2009},
and $\rho(S_n)$ is the
total level density calculated from equation~\ref{eq:rhoSn}. Finally, $\rho^{\mathrm{syst}}$ is the
total level density at $S_n$ as predicted from the global systematics of ~\cite{egidy2009}. $^\dagger$Estimated from systematics.}
\begin{tabular}{lcclcllll}
\hline 
\hline
A$_f$      & $J_t$     & $S_n$ & $D_0$      & $\sigma(S_n)$ & $Pa^{\prime}$ & $\rho(S_n)$         & $\rho^{\mathrm{syst}}(S_n)$ & $\left< \Gamma_{\gamma 0}\right>$   \\
&           & (MeV) &  (keV)     &               & (MeV)         &(10$^{3}$ MeV$^{-1}$)& (10$^{3}$ MeV$^{-1}$)       & (meV)                               \\
\hline
$^{55}$Fe  &  0        & 9.298 & 20.5(14)   & 3.41          & 0.463         & 1.19(9)             & 1.28                        & 1600(700)                           \\
$^{56}$Fe & 3/2& 11.197&3.36(124)$^\dagger$&3.47&2.905        & 2.18(59)$^\dagger$  & 2.94                        & 1900(600)$^\dagger$                 \\
$^{57}$Fe  &  0        & 9.298 & 25.4(22)   & 3.35          & 0.211         & 0.926(80)           & 1.14                        & 920(410)                           \\
$^{58}$Fe  & 1/2       & 10.044& 7.05(70)   & 3.44          & 2.874         & 1.81(18)            & 3.49                        & 1850(500)                           \\
$^{59}$Fe  & 0         & 6.581 & 21.6(26)   & 3.30          & 0.470         & 1.06(13)            & 1.01                        & 1130(110)                           \\
\hline
\hline
\end{tabular}

\label{table1}
\end{center}
\end{table*}

\begin{table*}[!h]\footnotesize
\begin{center}
\caption{Neutron resonance parameters $D_0$ from~\cite{RIPL3}, 
and spin cutoff parameters from global systematics of~\cite{egidy2005};
$A_f$ is the final nucleus following neutron capture, $J_t$ is the ground-state spin of the target nucleus, $S_n$ is the neutron-separation energy, 
$\sigma$ is the spin-cutoff parameter from equation~(\ref{eq:spincut05}), $D_0$ is the $s$-wave level spacing~\cite{RIPL3},
$a$ and $E_1$ are the level density parameter and energy shift from~\cite{egidy2005}, 
and $\rho(S_n)$ is the
total level density calculated from equation~\ref{eq:rhoSn}. Finally, $\rho^{\mathrm{syst}}$ is the
total level density at $S_n$ as predicted from the global systematics of ~\cite{egidy2005}. $^\dagger$Estimated from systematics.}
\begin{tabular}{lcclcclll}
\hline
\hline
A$_f$      & $J_t$     & $S_n$ & $D_0$      & $\sigma(S_n)$ & $a$    & $E_1$  & $\rho(S_n)$         & $\rho^{\mathrm{syst}}(S_n)$    \\
&           & (MeV) &  (keV)     &               & (1/MeV)& (MeV)  &(10$^{3}$ MeV$^{-1}$)& (10$^{3}$ MeV$^{-1}$)                                  \\
\hline
$^{55}$Fe  &  0        & 9.298 & 20.5(14)   & 4.02          & 5.817  & -0.524 & 1.62(11)            & 2.00                                    \\
$^{56}$Fe &3/2&11.197&3.30$^{+0.9}_{-0.6}$$^\dagger$&4.05&6.196&0.942& 2.87(68)$^\dagger$& 4.22                                    \\
$^{57}$Fe  &  0        & 9.298 & 25.4(22)   & 3.83          & 6.581  & -0.523 & 1.20(10)            & 1.62                                  \\
$^{58}$Fe  & 1/2       & 10.044& 7.05(70)   & 3.93          & 6.936  & 0.942  & 2.32(23)            & 4.66                                   \\
$^{59}$Fe  & 0         & 6.581 & 21.6(26)   & 3.70          & 7.297  & -0.424 & 1.32(16)            & 1.38                                  \\
\hline
\hline
\end{tabular}
\label{table2}
\end{center}
\end{table*}

\begin{table*}[!h]\footnotesize
\begin{center}
\caption{Parameters for the constant-temperature interpolation for the different normalization options. 
Both parameters $T$ and $E_0$ are given in MeV.}
\begin{tabular}{lrrrrrrrrrrrr}
\hline
\hline
&\multicolumn{6}{c}{Norm-1}    &  \multicolumn{6}{c}{Norm-2} \\
\hline
Nucleus   &\multicolumn{2}{c}{Lower} &\multicolumn{2}{c}{Middle} &\multicolumn{2}{c}{Upper} &\multicolumn{2}{c}{Lower} &\multicolumn{2}{c}{Middle} &\multicolumn{2}{c}{Upper} \\
          & $T$ & $E_0$    & $T$ & $E_0$     & $T$ & $E_0$     & $T$ & $E_0$     & $T$ & $E_0$    & $T$ & $E_0$    \\
\hline
$^{56}$Fe & 1.41  & 0.320  & 1.40  &$-0.034$ & 1.38  &$-0.169$ & 1.40  &$-0.070$ & 1.35  & 0.045  & 1.30  &0.232 \\                         
$^{57}$Fe & 1.32  &$-1.618$& 1.30  &$-1.575$ & 1.29  &$-1.601$ & 1.31  &$-1.882$ & 1.29  &$-1.829$& 1.28  &$-1.848$ \\                              
\hline
\hline
\end{tabular}
\label{tab:ctvalues}
\end{center}
\end{table*}
\newpage

\section{More details for the angular distributions}
\label{AppendixB}
For the Legendre polynomials we have
\begin{equation}
P_2(\cos \theta) = \frac{1}{2}\left[3(\cos\theta)^2 -1 \right],
\end{equation}
\begin{equation}
P_4(\cos \theta) = \frac{1}{8}\left[35(\cos\theta)^4 - 30(\cos\theta)^2 +3 \right].
\end{equation}

In the case of a fully aligned state with respect to the beam direction ($\alpha_k = 1$), 
the $a_k^{\mathrm{max}}$ coefficients are given by~\cite{mateosian1974}
\begin{equation}
a_k^{\mathrm{max}}(J_i L L^\prime J_f) = 
\frac{B_k}{1 + \delta^2} \left[F_k(J_f L L J_i) + 2\delta F_k(J_f L L^\prime J_i) 
+ \delta^2F_k(J_f L^\prime L^\prime J_i)\right].
\end{equation}
Here, $J_i, J_f$ are the spins of the initial and final level, $L, L^\prime$ are transition multipolarities, 
$\delta$ is the mixing ratio between the multipolarities defined according to~\cite{krane1978}:
\begin{equation}
\delta = \frac{ \left< J_f || E(L+1) || J_i \right> } {\left< J_f || M(L) || J_i \right>}.
\end{equation}
Here, $E(L+1)$ is the electric transition operator for multipolarity $L+1$, and $M(L)$ is the magnetic 
transition operator
for multipolarity $L$. Further,
the $B_k, F_k$ coefficients are defined in~\cite{mateosian1974}, where also values for the product 
$B_k F_k$ are tabulated.

We investigate known transitions in $^{57}$Fe, such as the 692-keV $\gamma$ ray decaying from the level
at $706$ keV, where $J_i = 5/2^-$ and $J_f = 3/2^-$, and the transition is known to be of $M1+E2$ type
with a mixing ratio $\delta = -0.465$~\cite{ensdf}. We get 
\begin{eqnarray*}
a_2^{\mathrm{max}} & = & \frac{1}{1+0.465^2} [B_2 F_2(3/2,1,1,5/2) + 2\cdot (-0.465) \cdot B_2 F_2(3/2,1,2,5/2) \\
& + & 0.465^2 B_2 F_2(3/2,2,2,5/2) ].
\end{eqnarray*}
From~\cite{mateosian1974} we have $B_2 F_2(3/2,1,1,5/2) = -0.400$, $B_2 F_2(3/2,1,2,5/2) = 1.014$, 
and $B_2 F_2(3/2,2,2,5/2) = 0.204$, giving $a_2^{\mathrm{max}} = -1.068$. For $a_4^{\mathrm{max}}$, 
we find
\begin{equation*}
a_4^{\mathrm{max}}  =  \frac{1}{1+0.465^2} [0.465^2 B_4 F_4(3/2,2,2,5/2)];
\end{equation*}
with $B_4 F_4(3/2,2,2,5/2) = 0.653$, we get $a_4^{\mathrm{max}} = 0.116$. 
Similarly, we get for an $E2$ transition with $J_i = 9/2, J_f = 5/2$ and no mixing ($\delta = 0$), 
$a_2^{\mathrm{max}} = 0.476$ and $a_4^{\mathrm{max}} = -0.286$. 

\end{document}